\documentclass[aps, twocolumn, prx,showpacs, longbibliographye]{revtex4-2}

\usepackage{epsfig}
\usepackage{color}
\usepackage{mhchem}
\usepackage[makeroom]{cancel} 
\usepackage[normalem]{ulem} 
\usepackage{xcolor}
\pagecolor{white}
\usepackage{amssymb,graphics,graphicx,amsfonts,amsmath,empheq}
\usepackage[breakable,most,many]{tcolorbox}
\usepackage{varwidth}
\usepackage{capt-of}
\usepackage[export]{adjustbox}
\usepackage{hyperref}

\begin{document}

\title{Liquid Hopfield model: retrieval and localization in multicomponent liquid mixtures}

\author{Rodrigo Braz Teixeira}
\affiliation{Gulbenkian Institute of Molecular Medicine, 2780-156 Oeiras }
\affiliation{Centro de Física Teórica e Computacional, Faculdade de Ciências, Universidade de Lisboa, 1749-016 Lisboa, Portugal }
\author{Giorgio Carugno}
\affiliation{Department of Mathematics, King’s College London, 
 Strand, London, WC2R 2LS, United Kingdom }
\author{Izaak Neri}
\affiliation{ Department of Mathematics, King’s College London, 
 Strand, London, WC2R 2LS, United Kingdom}
\author{Pablo Sartori}
\affiliation{Gulbenkian Institute of Molecular Medicine, 2780-156 Oeiras, Portugal}

\date{\today}

\begin{abstract}
{\bf Abstract.} Biological mixtures, such as the cellular cytoplasm, are composed of a large number of different components. From this heterogeneity, ordered mesoscopic structures emerge, such as liquid phases with controlled composition. These structures compete with each other for the same components. This raises several questions, such as what types of interactions allow the {\it retrieval} of multiple ordered mesoscopic structures, and what are the physical limitations for the retrieval of said structures.   In this work, we develop an analytically tractable model for liquids capable of retrieving states with target compositions. We name this model the {\it liquid Hopfield model} in reference to  corresponding work in the theory of associative neural networks. By solving this model, we show that non-linear repulsive interactions are a general requirement for retrieval of target structures. We demonstrate that this is because liquid mixtures at low temperatures tend to transition to phases with few components, a phenomenon that we term {\it localization}.   Taken together, our results demonstrate a trade-off between retrieval and localization phenomena in liquid mixtures.
\newline

{\bf Significance statement.} The cellular cytoplasm is a liquid mixture  of thousands of different components that interact in diverse ways. This mixture self-organizes into ordered structures, such as liquid phases with controlled composition. It remains unclear what types of interactions among components ensures the reliable assembly of these phases.  To answer this question, we establish an analogy between neural networks and multi-component liquid mixtures. This allows us to demonstrate that reliable assembly of functional liquid phases is at odds with the tendency of mixtures to segregate into regions enriched in few components.  This work unravels a trade-off between these types of phase behaviour and constitutes a first step towards linking the fields of neural networks to liquid mixtures.

\end{abstract}

\pacs{}
\maketitle

\begin{center}
\textbf{INTRODUCTION}
\end{center}

Within cells, thousands of different components, including proteins, nucleic acids and small peptides, interact with each other  \cite{sear2003instabilities, gavin2002functional}. From this heterogeneous mixture, mesoscopic scale structures are \textit{retrieved} to perform specific biological functions. Examples of such retrieval are the assembly of solid-like multi-protein complexes  \cite{marsh2015structure,gavin2002functional}, demixing of liquid droplets in the cytoplasm \cite{brangwynne2009germline, hyman2014liquid, choi2020physical, berry2018physical}, or formation of lipid rafts on the membrane \cite{lingwood2010lipid, heberle2011phase}. Concomitant with their biological function, the composition of such structures is ordered, in the sense that it is tightly controlled. Therefore, the parsimonious coexistence of heterogeneity and order is an essential characteristic of cellular mixtures.

Conventional statistical mechanics and soft matter physics, focuses on systems in which the number of component species, $N$, is much smaller than the total number of components per species, $M$.  The standard thermodynamic limit is thus $N\ll M$ and $M\to \infty$.  In contrast,  biological mixtures often take place in the regime $N,M\to\infty$, and exhibit phases in which the number of components enriched is also large, $Q\lesssim N$.  Considering the alternative thermodynamic limit with $N,M\to\infty$, elsewhere referred to as multifarious \cite{murugan2015multifarious, sartori2020lessons}, is key to   describe  the concurrence of heterogeneity and order in biological matter.

Previous studies of liquid mixtures with many components can be roughly grouped in two categories. The first, pioneered by Sear and Cuesta~\cite{sear2003instabilities}, uses analytical tools to study stability of the homogeneous state of a mixture with random interactions~\cite{sear2003instabilities, jacobs2013predicting, jacobs2017phase,  carugno2022instabilities, girard2022kinetics, sollich}. The second, uses numerical inverse optimization on the interactions so to enforce stability of multiple target phases \cite{zwicker2022evolved, jacobs2021self}. However, both approaches have important caveats: the stability properties of the homogeneous phase cannot be used to predict the properties of ordered phases, which are in any case unlikely for random interactions; and while inverse optimization guarantees by construction stability of target phases and provides numerical  values for the interaction matrices, these shed   little light on the  rules that govern the interactions that guarantee such stability. Due to these limitations, two central questions remain unanswered: what types of interactions ensure retrieval of target ordered structures, and what are the physical trade-offs of target retrieval in heterogeneous mixtures?

To answer these questions, we follow an alternative approach: we prescribe a set of interactions and analytically derive the conditions under which these guarantee stability of target phases. Our choice of interactions is based on an analogy between the nucleation of target liquid phases and the retrieval of patterns in Hopfield neural networks \cite{hopfield1982neural}, see also \cite{murugan2015multifarious, sartori2020lessons}. We thus refer to this model as the \textit{liquid Hopfield model}. For this model, we find conditions to ensure stability of up to $p_{\rm max}=N-1$ target phases, for arbitrary values of $N$ and $Q$. One key condition is the presence of non-linear repulsive interactions, which stabilize retrieval. We further show that when such non-linearities are weak the mixture tends towards localized phases, in line with prior numerical evidence \cite{jacobs2013predicting, jacobs2017phase}, and show that full localization occurs at zero temperature. Overall, this work establishes that multi-component mixtures exhibit a trade-off between localization and retrieval, with non-linearities playing a key role in leveraging this trade-off.

The paper is organized as follows. First, we discuss a generic thermodynamic model for multi-component liquid mixtures in the grand-canonical ensemble. Next, we formulate the problem of retrieval of target phases and define the affinity matrix. This completes the {\it liquid Hopfield model}. In the sections that follow, we study homogeneous, retrieval and localized states for a particular choice of targets. In the last two sections, we show how our results generalize to arbitrary targets. Lastly, we discuss the results of this paper and put them in the context of current advances in multi-component mixtures.

\vspace{.1in}

\begin{center}
\textbf{RESULTS}
\end{center}

{\bf Thermodynamics of heterogeneous fluid mixtures.}\label{sec:thermo} We consider a multi-component fluid mixture of $N$ molecular species with densities $\rho_i$, where $i=1,2,\ldots,N$. The thermodynamic behavior of this fluid  is dictated by the free energy density
\begin{equation}\label{eq:f}
f(\vec{\rho},T) = u(\vec{\rho}) - T \: s(\vec{\rho}), 
\end{equation}
where $u$ is the  energy density, $s$ is the entropy density  \cite{callen1998thermodynamics}, and $T$ is the temperature; 

A generic form for the entropy density can be obtained by adding the entropy of each of the components to that of the solvent, see e.g.~Refs.~\cite{samsafransbook, doi2013soft, jacobs2013predicting, mao2019phase}. The resulting expression is
\begin{equation}\label{eq:s}
s(\vec{\rho}) = -k_{\rm B}\sum^N_{i=1}\rho_i\log(\rho_i)-k_{\rm B}(1-\rho)\log(1-\rho),
\end{equation}
where $k_{\rm B}$ is Boltzmann's constant (in what follows we set $k_{\rm B}T=1$) and $\rho=\sum_{i=1}^N\rho_i$ the total density. Nota bene, the gradient of the entropy diverges for $\rho_i=0$ or $\rho=1$, which constrains the densities to the interior of the $N-1$ dimensional standard simplex.

We characterize the internal energy via a low-density expansion and truncate to cubic order, instead of the common quadratic form \cite{sear2003instabilities,jacobs2013predicting}.  For simplicity, we take the cubic term to be diagonal, and so
\begin{align}\label{eq:u}
u(\vec{\rho}) =-\frac{v_2}{2} \sum^N_{i=1}\sum^N_{j=1}J_{ij}\rho_i\rho_j+\frac{v_3}{6}\sum_{i=1}^{N}\rho_i^3N^{2}, 
\end{align}
where $v_2>0$ and $v_3>0$ are constants that quantify the strength of the interactions and $J_{ij}$ is the pairwise affinity matrix (the $N^2$ factor keeps the scale of the term proportional to $v_3$ independent of $N$ for $\rho_i\sim N^{-1}$). Note that the quadratic term can be attractive ($J_{ij}>0$) and repulsive ($J_{ij}<0$), whereas we only consider a repulsive  cubic term.   In the context of biological liquid mixtures,  cubic and higher order interactions can be expected due to the polymeric nature of proteins \cite{rubinstein2003polymer} and prevalence of allosteric effects \cite{phillips2020molecular, luo2024beyond}. The role of the cubic term in the present model will become clear later.

\begin{figure}[h]
\centerline{\includegraphics[]{ 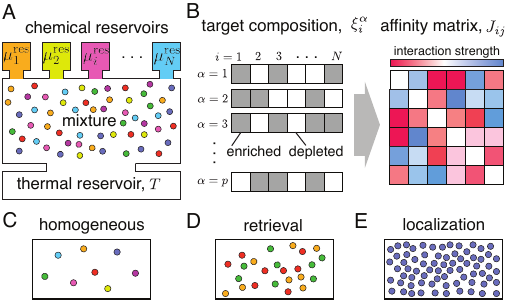}}
\caption{\label{fig:schema}{\it Schematic representation of the problem setup.} {\bf A.} A liquid mixture with components $i=1,\ldots,N$ is in contact with a thermal reservoir at temperature $T$ and chemical reservoirs with chemical potentials $\mu^{\rm res}_i$.   {\bf B.}  Binary composition target vectors $\xi_i^\alpha$ indicate which components should be enriched or depleted in the retrieval states of interest. The $p$ targets are used to design the affinity matrix, $J_{ij}$. {\bf C-E.} Depending on the thermodynamic parameters, we distinguish three types of  stable liquid states:  homogeneous states, where all components mix equally;  retrieval states, which are enriched/depleted according to the targets; and localised states, enriched in a small number of  components. } \label{fig:sketch}
\end{figure}

To study metastable states of this mixture, we consider a setting in which the fluid is in contact with a thermal reservoir at temperature $T$ and $N$ chemical reservoirs at chemical potentials  $\mu_i^{\rm res}$, see Fig.~\ref{fig:sketch}A. Hereafter, we assume for simplicity  that all chemical potentials are equal, i.e., $\mu_i^{\rm res}=\mu^{\rm res}$.   Metastable states are  determined by  densities $\vec{\rho}_\ast$ that satisfy the following two conditions.

The first condition is {\it chemical equilibrium}, by which the chemical potentials of the fluid, $\partial f/\partial{\rho_i}$, match those of the reservoir, i.e.,
\begin{align}\label{eq:locmin}
\frac{\partial f}{\partial{\rho_i}} (\vec{\rho}_\ast)=\mu^{\rm res},
\end{align}
for all $i\in \left\{1,2,\ldots,N\right\}$.

The second condition is {\it mechanical stability}, i.e.,  the Hessian $H_{ij}=\partial^2 f/\partial\rho_i\partial\rho_j$ (inversely proportional to the isothermal compressibility \cite{callen1998thermodynamics}) is positive semi-definite,
\begin{align}\label{eq:mech}
\sum^N_{i=1}\sum^N_{j=1} H_{ij}^\ast x_{i} x_j \ge0,
\end{align}
where $H_{ij}^\ast=H_{ij}(\vec{\rho}_\ast)$ and $\vec{x}$ are arbitrary vectors in $\mathbb{R}^N$.  The equality in (\ref{eq:mech}) is attained  when $\vec{\rho}_\ast$ is located at the \textit{spinodal} manifold.

Defining the  grand-potential functional
\begin{equation}
 \omega(\vec{\rho},\mu^{\rm res}) = f(\vec{\rho}) - \rho\mu^{\rm res}, \label{eq:omega}
 \end{equation}
chemical equilibria are stationary points of $\omega$ w.r.t.~$\vec{\rho}$ at fixed $\mu^{\rm res}$. Such points are mechanically stable when they are {\it local minima} of the grand-potential functional. 
\newline

{\bf Metastable targets and retrieval.}  \label{sec:meta} We use the framework above, to study the physical limitations on the stability of states with specified compositions.  Our goal is to construct an affinity matrix $J_{ij}$, so that the liquid exhibits metastable states corresponding to $p$ pre-defined {\it target} composition vectors $\vec{\xi}^\alpha$, with $\alpha=1,\ldots,p$. The {\it targets} have binary entries that determine whether component $i$ should be enriched ($\xi_i^\alpha=1$) or depleted ($\xi_i^\alpha=0$), see Fig.~\ref{fig:schema}A and B. We say that a fluid is then capable of retrieving the target state $\alpha$ if the corresponding metastable state is enriched as prescribed by $\vec{\xi}^{\alpha}$.

To endow the fluid with this retrieval capability, we draw inspiration from the classical work on neural networks by J.J. Hopfield \cite{hertz2018introduction} and propose as affinity matrix
\begin{align}\label{eq:Jij}
J_{ij} = \sum_{\alpha,\beta=1}^p\gamma_i^{\alpha}c_{\alpha\beta}^{-1}\gamma_j^{\beta},
\end{align}
where $\vec{\gamma}^{\alpha}= (\vec{\xi}^{\alpha} - q)/n$, $c_{\alpha\beta}=\sum_{i=1}^N\gamma_i^{\alpha}\gamma_i^{\beta}/N$ is a covariance matrix, and $c^{-1}_{\alpha\beta}$ its inverse. Here, $q=Q/N$ is the sparsity parameter, equal for all targets, with $Q$ the number of components enriched in the target compositions (i.e., for which $\xi^{\alpha}_i=1$, see Fig.~\ref{fig:schema}B), and $n = \sqrt{q\left(1 - q\right)}$ is a normalization factor. Due to its combination of concepts from multi-components fluids and neural networks, we refer to this model as the \textit{liquid Hopfield model}. Notice that for orthogonal $\vec{\gamma}^{\alpha}$ we recover
\begin{align}\label{eq:Jij_simple}
J_{ij} = \sum_{\alpha=1}^p\gamma_i^{\alpha}\gamma_j^{\alpha},
\end{align}
which we recognize as the interaction matrix used by Hopfield in Ref.~\cite{hopfield1982neural}.

Our approach to study retrieval in the liquid Hopfield model is as follows. We  first focus on a special set of targets, for which $\vec{\gamma}^{\alpha}$ are the columns of Sylvester-Hadamard matrices (see \textit{Materials and Methods} for a definition). We call these the \textit{hypersymmetric} targets. While a seemingly peculiar choice, we  later show that hypersymmetric targets are the most stringent choice, i.e., when these targets are stable, then  for the same model parameters any other set of targets are also stable. Therefore, we first focus on hypersymmetric targets and then generalize our results for arbitrary targets.
\newline

\begin{figure}[h]
\centerline{\includegraphics[]{ 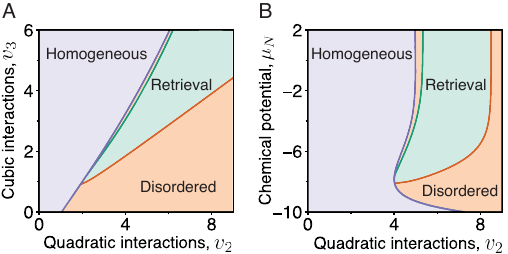}}
\caption{\label{fig:phase_diag}{\it Stability diagrams for hypersymmetric targets.} {\bf A.} Spinodal lines delimiting the regions where either the  homogeneous state or  all retrieval states are stable as a function of $v_2$ and $v_3$ with $\mu_N=-6$.   The plotted lines are given by the equalities in Eqs.~(\ref{eq:crit_temp_homogeneous}), (\ref{eq:stability_ansatz}) and (\ref{eq:stability_ansatz2}).  Note that  retrieval only occurs for large enough cubic interactions $v_3$.   
{\bf B.} Same as Panel  A, albeit as a function of $v_2$ and $\mu_N$ and with $v_3=4$. 
Note that  retrieval only occurs for large enough chemical potential   $\mu_N$. }
\end{figure}

{\bf State stability diagram.}\label{sec:stab} Figure~\ref{fig:phase_diag} shows the stability diagram of the liquid Hopfield model for hypersymmetric targets. The Figure depicts the region where the homogeneous state of the mixture is stable (violet), the region  where all $p$ targets can be retrieved (green), and two regions where retrieval is not possible (orange). Boundaries of these regions are the \textit{spinodal} lines of the corresponding states (global stability lines show similar behavior, see section~\ref{app:binodal} of the Supplementary Information, SI).

The stability of the homogeneous state for $v_2\approx0$ is expected, as in this regime contributions from entropy and cubic repulsion dominate, which mix all different components equally.

For sufficiently large repulsion, $v_3$, retrieval of target compositions is possible, and corresponding \textit{retrieval} states are stable. Between the homogeneous and retrieval region there is a small transition region. The retrieval region  expands as $v_2$ and $v_3$ increase.

For $v_2$ large enough, stability of retrieval states is lost and the system enters a disordered region in which many different non-retrieval states are locally stable. For $v_2\gg v_3$, states with few components, which we call \textit{localized} states,  are globally stable. As we clarify later on, since the cubic repulsive term penalizes such localized states, this elucidates the role of $v_3$  in stabilizing retrieval.
\newline

{\bf Homogeneous region.}\label{sec:hom} A homogeneous state of the mixture  has densities $\rho_i= \rho/N$ for all components $i$. The homogeneous states for hypersymmetric targets satisfy the condition of chemical equilibrium, given by Eq.~(\ref{eq:locmin}), when $\rho=\rho_\ast$ with
 \begin{equation}\label{eq:crit_point_homogeneous}
\rho_{\ast} = \frac{1}{1 +  \exp\left( - \mu_N +  \frac{v_3}{2}  \rho_{\ast}^{2}\right)},   
\end{equation}
and where  $\mu_N=\mu^{\rm res} + \ln N$ is a rescaled chemical potential  (Sec. \ref{app:chemeq} of the SI). Note that in the limit $N\gg1$ for fixed $\mu^{\rm res}$ it holds that  $\mu_N\gg1$ and $\rho_{\ast}\to1$, whereas for fixed $\mu_N$ we have that $\rho_{\ast}$ is independent of $N$. In the following, we  work on the  second, arguably more physical, ensemble.

Now, we study the stability of the homogeneous state.  Equation~(\ref{eq:mech}) holds if and only if all eigenvalues of the Hessian $H_{ij}$ are non-negative. We therefore diagonalize $H_{ij}$ at the homogeneous state analytically, details in \textit{SI}~\ref{app:homo_hess}. The non-negativity of the smallest eigenvalue, see \textit{Materials and Methods} for an explicit expression, yields the inequality
\begin{equation}\label{eq:crit_temp_homogeneous}
 \rho_{\ast} \left(v_2 -  v_3 \rho_{\ast} \right) \leq 1  .
\end{equation}
The equality in (\ref{eq:crit_temp_homogeneous})  is attained at the {\it spinodal} manifold, denoted by the violet lines in Fig.~\ref{fig:phase_diag}, for which the homogeneous state is marginally stable; in the low-temperature and dense limit, $v_2\gg1$ and $\mu_N\gg1$, this reduces to $v_3=v_2$. At the {\it spinodal} manifold, the unstable modes are spanned by the subspace of the target compositions $\vec{\gamma}^\alpha$. While this could be taken to suggest that the corresponding retrieval states are stable, see analysis in \cite{sear2003instabilities, brenner, carugno2022instabilities, machta}, we will later show that this is only the case in presence of cubic repulsion.
\newline

{\bf Retrieval region.} In the retrieval region, the mixture can adopt a stable density pattern according to a particular target $\alpha$. In particular, the density of component $i$ is  enriched  when  $\gamma^{\alpha}_i=1$ and depleted when $\gamma^{\alpha}_i=-1$. Therefore, to characterize retrieval we  use the  {\it ansatz} 
\begin{align}\label{eq:ansatz}
\rho_i^{\alpha} =\frac{\rho}{N}(1+a\gamma_i^{\alpha})
\end{align}
for retrieval states, where $a\in[-1,1]$ measures the degree of retrieval, and which we call the {\it overlap} between the state of the mixture  and the target state \cite{hertz2018introduction}. Imposing chemical equilibrium on this {\it ansatz} we find that retrieval states are chemically stable when $a=a_\ast$ and $\rho=\rho_\ast$ with
\begin{align}
a_{\ast} &= \tanh\left[  a_{\ast} \rho_\ast \left(v_2- v_3\rho_\ast\right)\right],\label{eq:rho_a_minA}\\
\rho_{\ast} &= \frac{1}{1 + \exp\left(-  \mu_N +  \frac{v_3}{2}  \rho_{\ast}^2 \left(1 +  a_{\ast}^2 \right) + \frac12\ln(1-a_{\ast}^2) \right)},\label{eq:rho_a_minB}
\end{align}
see Sec. \ref{app:chemeq} of the SI for a detailed derivation.  Note that $1>\rho_i^{\alpha}>0$ and $\rho^\alpha_i$ approaches $0$ for large values  of 
$v_2$. 

\begin{figure}[h]
\centerline{\includegraphics[]{ 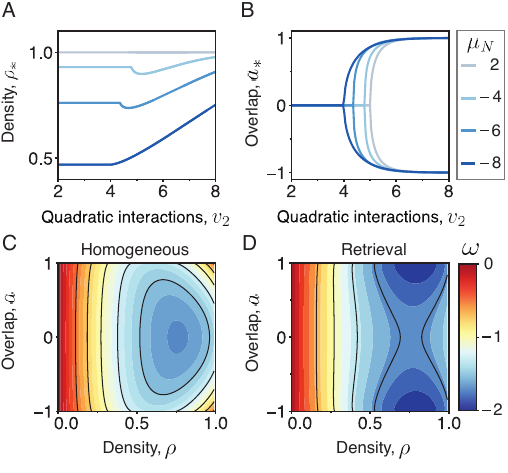}}
\caption{\label{fig:transition}{\it Chemical equilibrium of  homogeneous and retrieval states.} {\bf A. } The total density   $\rho_\ast$ as a function of the quadratic interaction strength $v_2$ for states in chemical equilibrium  is shown for different values of the chemical potential $\mu_N$ [Eq.~(\ref{eq:rho_a_minB})]. {\bf B. }  Same as Panel A, but for the overlap $a_\ast$ [Eq.~(\ref{eq:rho_a_minA})].    The overlap becomes non-zero outside the homogeneous region, indicating the emergence of the retrieval region.  {\bf C. } Heatmap of the   grand-potential functional $\omega$ evaluated at the retrieval ansatz (\ref{eq:ansatz}) as a function of $(a,\rho)$  for $v_2 = 2$,  $\mu_N = -6$, and $v_3 = 4$.  The minimum represents the homogeneous state. {\bf D. }  Same as Panel C but with $v_2=6$.  In this case there are two mirror symmetric minima ($a\leftrightarrow -a$) that correspond to retrieval states.}
\end{figure}

In Fig.~\ref{fig:transition}A and B, we plot the solutions of \eqref{eq:rho_a_minA} and \eqref{eq:rho_a_minB}  as a function of the quadratic interaction strength, $v_2$, for different values of the rescaled chemical potential, $\mu_N$. The figure shows that the retrieval state emerges from the homogeneous state at its spinodal line. We remark that due to the $ a_{\ast} \to - a_{\ast}$ symmetry in \eqref{eq:rho_a_minA} and \eqref{eq:rho_a_minB} (consequence of $q = 1/2$) a second ``mirror'' state also emerges in which enrichment is opposite to that of the target $\alpha$. Evaluating the   grand-potential functional $\omega(\vec{\rho})$ in the subspace spanned by Eq.~\ref{eq:ansatz} suggests that the retrieval state is not only a stationary point, but a stable minimum, see Fig.~\ref{fig:transition}C and D. However, a complete stability picture requires studying the $N$-dimensional stability condition in \eqref{eq:mech}, which  we do in the following.

Deriving an analytical expression for the smallest eigenvalue of the Hessian (see \textit{Materials and Methods} for the expression and SI ~\ref{app:min_eig} for a derivation), we find that a retrieval state is stable when the following two conditions hold
\begin{align}
    \rho_{\ast} \left(1 +  a_{\ast} \right) \left( v_2- v_3 \rho_{\ast} \left(1 +  a_{\ast} \right) \right)  &\leq  1\, ,\label{eq:stability_ansatz}\\ 
    \rho_{\ast} \left(1 - a_{\ast} \right) \left( v_2-  v_3 \rho_{\ast} \left(1 -  a_{\ast} \right) \right)  &\leq  1 \, ,\label{eq:stability_ansatz2}
\end{align}
where $(\rho_{\ast}, a_{\ast})$ are taken from \eqref{eq:rho_a_minA} and \eqref{eq:rho_a_minB}. The two equalities above correspond to the orange and green lines, respectively, of Fig.~\ref{fig:phase_diag} when $a_\ast>0$ (and the other way around when $a_\ast<0$).  The corresponding  instability modes are confined into the space spanned by the target compositions. Note that between the homogeneous and retrieval regions there exists a narrow transition band where neither the homogeneous state nor the retrieval states are stable. In the low-temperature and high density regime, the width of  this transition region scales asymptotically as $\log(v_2)$, and the width of the retrieval region as $v_2/2$. 

Two important facts can be deduced from \eqref{eq:stability_ansatz} and \eqref{eq:stability_ansatz2}. First, in absence of the repulsive cubic  interaction, i.e., $v_3=0$, adding the inequalities gives $\rho_\ast v_2\le1$, which by Eq.~\eqref{eq:rho_a_minA} implies that $a_\ast=0$, and so there are no retrieval states, see also Fig.~\ref{fig:phase_diag}. Therefore, sufficiently strong non-linear repulsion is necessary for retrieval. Second, Eq.~\eqref{eq:stability_ansatz} does not depend on the number of species $N$ nor on the number of targets $p$. Consequently, the hypersymmetric liquid Hopfield model allows up to $p_{\rm max}=N-1$ target compositions to be simultaneously stable, provided cubic repulsion is strong enough.

\begin{figure}[h]
\centerline{\includegraphics[]{ 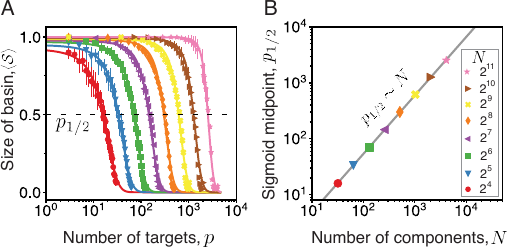}}

\caption{\label{fig:scaling}{\it Scaling of basins of attraction of retrieval states} {\bf A. } The mean size of the basin of attraction of retrieval states, $\langle\mathcal{S}\rangle$, is plotted against the number of targets, $p$, for mixtures with different number of components, $N$. Sigmoidal curves are fit to each set of points. As the number of targets increases, the size of the basin of attraction shrinks. {\bf B. } Scaling of the mid-point of each sigmoidal, $p_{1/2}$, as a function of the number of components in the mixture, $N$. A linear scaling is compatible with the observed trend (line corresponds to $p_{1/2} = (0.622\pm0.002) N -(11\pm3)$). In both panels, hypersymmetric targets are considered and the following parameters remain fixed: $\mu_N = 1$, $v_2 = 6$ and $v_3 = 4$. See Sec.~\ref{app:methods} of the SI for a detailed description on the approach followed to produce this figure.}
\end{figure}

So far, we have shown that stability of retrieval states is independent on the number of components, $N$, and hypersymmetric targets, $p$. We now determine how the sizes of the basins of attraction depend on $p$ and $N$. To this end, we quantify the size of a basin of attraction in terms of the number of component densities in an initial state that need to be set equal to corresponding values in the retrieval state of interest in order for a gradient descent dynamics in $\omega$ to converge to that retrieval state. While gradient descent does not capture the dynamics of phase separation, in the grand-canonical setting it does characterize the response of the mixture to a small perturbation. We denote the associated observable as $\mathcal{S}$ (see  Sec.~\ref{app:methods} of the SI for a definition).

Figure \ref{fig:scaling}A, shows how the average size of the basins, $\langle\mathcal{S}\rangle$, depends on the number of targets, $p$, for mixtures with different number of components, $N$. Basins decrease in size as $p$ increases, as expected, and so  retrieval is effectively more difficult as more targets are encoded. The mid-point of this decrease, $p_{1/2}$, measures the capacity of a liquid mixture to encode different target compositions. Interestingly, we find that this mid-point increases linearly with the number of components, $N$, see Figure \ref{fig:scaling}B. This implies that as the number of components in a mixture increases, the number of states that can be retrieved increases proportionally.  Note that a similar scaling behaviour of the basins of attraction can be observed in associative neural networks \cite{hertz2018introduction}. 
\newline

{\bf Localization in the disordered region.} \label{sec:localization} We now focus on the disordered region, where neither the homogeneous nor the retrieval states are stable. 
Instead, in this region the stable states of the liquid mixture overlap with multiple target compositions.  Since for such states the liquid mixture develops into a "pathological" combination of the prescribed target states, we call these states non-retrieval states.   If the non-retrieval state is enriched in a small fraction of components then we say that the state is localised.  

We first consider the analytically solvable case of $v_3=0$, $\mu^{\rm res} = 0$, and $v_2\gg 1$. In this case, states in which only one component is enriched, i.e., $\rho_i = \delta_{i,k}$  with $k\in \left\{1,2,\ldots,N\right\}$ fixed, are minima of the  grand-potential functional $\omega(\vec{\rho})$ located at the corners of the simplex. Furthermore, for large enough values of  $p$ these states are the only configurations that minimise $\omega$, as shown in Sec.~\ref{app:zeroT} of the SI.  Corners of the simplex are fully localised states in the set of components, in the sense that they are enriched in only one component, and reminiscent of Anderson localisation for electrons in disordered materials~\cite{kramer1993localization, efetov1999supersymmetry}. Therefore, for the case considered and $p$ large, minima in the disordered region correspond to localized states.

\begin{figure}[h]
\centerline{\includegraphics[]{ 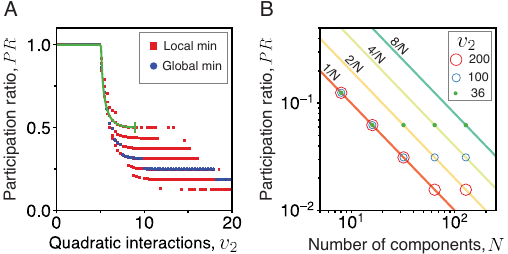}}
\caption{\label{fig:loc}{{\it Non-retrieval and localization.}  {\bf A. } Participation ratio, $PR$, of local and global minima of the grand-potential functional $\omega$ as a function of the quadratic interaction strength, $v_2$ for $N = 32$, $p = 17$, $v_3 = 6$ and $\mu_N=1$.  The green line indicates the participation ratio  $1/(1+a^2_\ast)$ of the homogeneous and retrieval states.  Observe how the $PR$ gradually decreases as a function of $v_2$ which is a signature of localisation in the limit of large $v_2$. {\bf B. }  Finite size scaling analysis of the participation ratio of the global minimum for given values of $v_2$ and with $p=24$, $\mu_N=1$, and $v_3=3$.   Note that the saturation value at large $N$ decreases as a function of $v_2$. Markers represent the $PR$ of states obtained from  numerically minimizing the  grand-potential functional $\omega(\vec{\rho})$ for hypersymmetric targets, as explained in  Sec.~\ref{app:methods} of the SI}.  }
\end{figure}

More generally, localization is quantified by the {\it participation ratio} 
\begin{align}\label{participation_ratio}
{\rm PR}(\left\{\rho_i\right\}) = \frac{1}{N}\frac{\left(\sum^N_{i=1}\rho_i\right)^2}{\sum^N_{i=1} \rho_i^2} ,
\end{align} 
with $\lim_{N\rightarrow \infty}{\rm PR}(\left\{\rho_i\right\})>0$ for a delocalised state, and $\lim_{N\rightarrow \infty}{\rm PR}(\left\{\rho_i\right\})=0$ for a localised state. Figure~\ref{fig:loc}A shows the participation ratio as a function of $v_2$ of minima of $\omega$ obtained numerically. For low $v_2$, the homogeneous state is stable, and so ${\rm PR}=1$. At intermediate $v_2$, the retrieval state becomes stable, for which ${\rm PR}=1/({1+a_{\ast}^2})$. In addition, we also observe other non-retrieval states that are locally stable, which are reminiscent of spurious states in associative neural networks \cite{hertz2018introduction}. For large values of $v_2$, these non-retrieval states have a decreasing PR, with the discrete jumps observed corresponding to the full depletion of components one by one. 

We further studied the role of increasing the number of components on localization. Fig.~\ref{fig:loc}B shows that the PR saturates to a constant value as $N$ increases. This asymptotic value decreases as a function of the quadratic interaction, $v_2$, and the number of targets, $p$, see  Sec.~\ref{app:zeroT} of the SI. Therefore, stable states are localized for $v_2\gg1$ and an extensive number of targets $p\sim N$, in agreement with the analytical results at the beginning of this section. 

Taken together, stable states in the disordered region have a small PR and localize in the limit of $v_2$ large and many targets $p\sim N$. This supports the idea that stability of the retrieval region is related to the suppression of localization.
\newline

\begin{figure}[h]
\centerline{\includegraphics[]{ 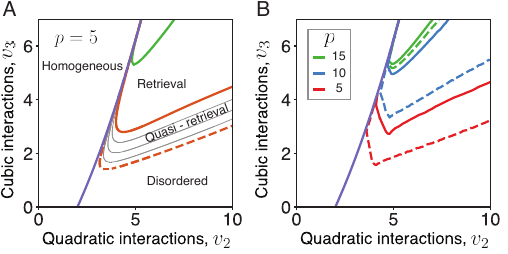}}\caption{\label{fig:Jijmod}{\it Stability diagrams for generic targets and $q=1/2$.} {\bf A.} Spinodal lines delimiting the regions where  the homogeneous state or each of the $p=5$ retrieval states are stable as a function of $v_2$ and $v_3$. Note that there are $p$ distinct spinodal lines, one for each of the target states, which yields the quasi-retrieval region. The stability line of retrieval for hypersymmetric targets is shown in green. {\bf B.} Stability diagrams for different number of target states $p$. Lines are obtained by averaging the spinodal lines at each value of $v_3$ for fifty sets of targets independently generated. The solid and dashed lines of each color delimit the retrieval and quasi-retrieval region for the corresponding $p$.  As $p$ increases, lines delimiting the quasi-retrieval and retrieval region collapse  on the retrieval spinodal for hypersymmetric targets. Targets have been generated through a random permutation of a vector with $q=1/2$. Parameters are fixed for both panels to $\mu_N = 0$ and $N = 32$. Spinodal lines for generic targets are obtained numerically by diagonalizing the hessian at states in chemical equilibrium, see Eqs.~(\ref{eq:chemEqQ_MM}-\ref{eq:chemEqRhoQ_MM}) }
\end{figure}

{\bf Generalization to any type of targets.}\label{sec:gen} Up until now, we have restricted our study to the case of hypersymmetric targets. We now generalize the results to completely generic targets. First, we will discuss the case of $q=1/2$, for which the stability diagram appears in Fig.~\ref{fig:Jijmod}.

For generic targets,  the characteristics of the homogeneous state remain unchanged, so that Eqs.\eqref{eq:crit_point_homogeneous} and \eqref{eq:crit_temp_homogeneous} apply (see  Sec.~\ref{app:homo_hess} of the SI). Furthermore, the chemical equilibrium condition of the retrieval states is also valid, and so Eqs.~\eqref{eq:rho_a_minA} and 
\eqref{eq:rho_a_minB} generalize to generic targets (see Sec.~\ref{app:chemeq} of the SI). Moreover, for generic targets   the retrieval region is larger   than for hypersymmetric targets, and thus the latter represent a worse-case-scenario for retrieval. In particular, in  Sec.~\ref{app:min_eig} of the SI we mathematically prove that the conditions in Eqs.~\eqref{eq:stability_ansatz} and \eqref{eq:stability_ansatz2} are sufficient conditions for retrieval. Therefore, retrieval is stable in a region at least as large as in the case of hypersymmetric targets.  See the green line of Fig.~\ref{fig:Jijmod}A. which corresponds with the spinodal line for the retrieval region in the case of hypersymmetric targets.

Beyond this, we find that each retrieval state has a distinctive \textit{spinodal} line. This results in a new region of parameters for which only some of the retrieval states are stable, which we term \textit{quasi-retrieval}, see Fig.~\ref{fig:Jijmod}A. Interestingly, Fig.~\ref{fig:Jijmod}B reveals that as the number of targets increases the quasi-retrieval region gradually decreases in size until it coincides with the \textit{spinodal} line of hypersymmetric targets, see  Sec.~\ref{app:min_eig} of the SI for a proof.   Hence, for a large number of targets $p$, we recover the worse-case-scenario for hypersymmetric targets.
\newline

{\bf Effect of sparsity on retrieval.} So far, we have restricted our study to generic targets for which $q=1/2$. In this case half of the components are present in each target.  We now consider the case of lower values of $q$, where less components are shared among targets.

\begin{figure}[h]
\centerline{\includegraphics[]{ 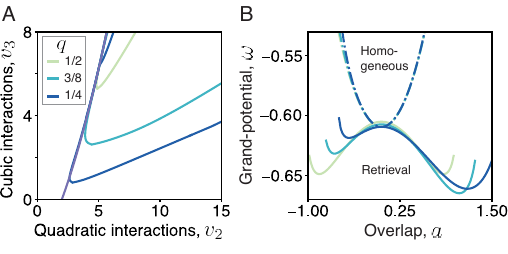}}\caption{\label{fig:sparse}{\it Stability diagrams for targets with variable sparsity.} {\bf A.} Lines delimit the region where the homogeneous state is locally stable and regions where all possible retrieval states with a given $q$ are  guaranteed to be locally stable [see  Eqs.~(\ref{eq:1Q}) and (\ref{eq:2Q}) in Sec.~\ref{app:min_eig} of the SI].  Parameters chosen are $p=25$,  $N = 32$, and $\mu_N = 0$.    {\bf B.} The   grand-potential functional $\omega$ evaluated at the retrieval ansatz  (\ref{eq:ansatz}) with $\rho=\rho_\ast$  is plotted as a function of  the overlap $a$,   where $\rho_\ast$ solves Eq.~(\ref{eq:chemEqRhoQ_MM}) for a given $a$.  Different colours correspond with   varying values of the sparsity $q$.  The single-well potential corresponds to the homogeneous regime $v_2 = 2$ and the double-well potential to the retrieval $v_2 = 8.5$.  Other parameters chosen are $v_3=4$, $\mu_N =0 $. Notice that in the retrieval region the  mirror symmetry  $a \to -a$ is broken for $q\neq 1/2$.  }
\end{figure}

Fig.~\ref{fig:sparse}A shows the region where the homogeneous state is stable and regions where all retrieval states are  also stable for different values of the sparsity, $q$.    Note that the spinodal for the homogeneous state, which we derive analytically in Sec.~\ref{app:homo_hess} of the SI, is independent of the sparsity $q$ of the target states.    In contrast with the homogeneous state, we find that stability of target states depends on the sparsity, and in fact reducing $q$, results in a larger retrieval region, consistent with previous results on multifarious self-assembly \cite{sartori2020lessons} and classical results of neural networks \cite{tsodyks1988enhanced}. Specifically, the lines plotted in  Fig.~\ref{fig:sparse}A represent the worse case scenario for the stability of target states in the large $p$ limit, and we derive those analytically in  Sec.~\ref{app:chemeq} of the SI.     Interestingly, we find that the key result that nonlinearities are necessary for retrieval, i.e. retrieval requires $v_3\neq0$, still hold even in the limit $q\to0$, as shown in  Sec.~\ref{app:sparse} of the SI.   

Recall that for $q=1/2$  each target state came accompanied by a mirror state for which $a_\ast<0$.    
Changing the sparsity of the targets from $q = 1/2$ breaks the symmetry $a_\ast\to-a_\ast$ in the corresponding equations. As a consequence, the values of the grand-potential functional for the mirror states increases as $q$ decreases, see Fig.~\ref{fig:sparse}B. Hence, the case of low $q$ thermodynamically suppresses mirror states, as desired.

Taken together, this shows that sparse targets facilitate retrieval, by increasing the parameter region where retrieval is stable and by suppressing potentially undesired mirror states.
\vspace{.1in}

\begin{center}
\textbf{DISCUSSION}
\end{center}
\label{sec:disc}
{\bf Summary.} Biological mixtures often consist of a large number of different components \cite{simons2000lipid, lingwood2010lipid, chong2016liquid, shin2017liquid, banani2017biomolecular}. Despite this, such mixtures are not disordered, but instead assemble into functional states with well characterized composition. Resolving this apparent paradox is challenging because there is no established physics formalism to describe materials with such characteristics. 

In this paper, we have introduced the liquid Hopfield model as a simple paradigm for multi-component mixtures capable of retrieving multiple target compositions. Our analysis revealed a trade-off between retrieval of target states, related to functional biological states; and disordered localized states, in which only few components are enriched.   As we have shown in this paper, the trade-off can be controlled by  repulsive non-linear interactions that penalize localisation, and which can be interpreted as  higher order terms in the virial expansion of the free energy.     
\newline

{\bf Trade-off between localisation and retrieval.}   
The retrieval capability of heterogeneous mixtures in the liquid Hopfield model is limited by the tendency for these mixtures  to localise at low temperatures.  Recall that we say a state is localised if in the large $N$ limit it is enriched in a finite number of components.    The relevance of localisation on the stability of states in heterogeneous mixtures is  put into evidence  through, among others,  the requirement of a nonlinear cubic repulsion term in the energy function.    Here, we further discuss the generality of this result that  captures a fundamental feature of   the physics of heterogeneous mixtures, such as those that occur in biological fluids.   

Localisation in heterogeneous mixtures is a generic phenomenon.    In fact, using a    zero temperature  argument,  we show in section~\ref{app:posdef} of the SI material that localisation occurs generically in heterogeneous liquid mixtures with a positive definite pairwise affinity matrix $J_{ij}$.  If $J_{ij}$ has negative eigenvalues, then the zero temperature minimisation problem is an NP-hard problem~\cite{murty1985some, pardalos1988checking,  pardalos1991quadratic},  which implies a rugged low-temperature free-energy landscape akin to a spin glass phase~\cite{mezard1987spin, charbonneau2023spin}. Hence, localisation is not specific to the interactions considered in this paper, and instead holds for generic interaction matrices.  
  
Given that localisation is a generic feature of heterogeneous liquid mixtures for large values of $N$, it also  follows that retrieval requires, in general, a nonlinear  repulsion.   Indeed,  nonlinear repulsion is necessary to confine the densities away from the boundaries of the $\vec{\rho}$-simplex,  which suppresses localization and enables retrieval. We remark that such confining non-linearity is needed despite the presence of nonlinear entropic terms, which are sufficient to provide chemical stability, but not mechanical stability, unlike for binary mixtures.   There is however no need to consider the specific choice of nonlinear term  considered in the present paper, and in section~\ref{app:alt-nonlin} of the SI we show that this paper's main results are preserved with other type of nonlinear interaction terms.  In the main  text we have, for simplicity,    considered the  case of a (diagonal) cubic interaction.  The need for ``strong’’ nonlinearities is in agreement with results of neural network models for continuous neuronal activity. In fact, the classic continuous implementation of J.J.~Hopfield ~\cite{hopfield1984neurons} required a strong sigmoidal nonlinearity for retrieval. Nonlinearities are also key in more modern variants of Hopfield networks with continuous variables~\cite{bolle2003spherical, Mcgraw1, Mcgraw2}, which require a  quartic term in the energy functional for successful retrieval,  as   otherwise the free energy functional  does  not have a number of minima that scales with $N$, see the Review~\cite{castellani2005spin} on spin glass theory.
  
We note that the trade-off between retrieval and localisation is a feature of heterogeneous mixtures with targets that are enriched in many components, i.e., $Q\sim N$ and $N$ large (or alternatively, $q>0$).  If the target states are themselves localised, corresponding with the case for which $Q$ is finite and $q=0$,  then retrieval is possible without nonlinear repulsion, as shown in Ref.~\cite{jacobs2021self}. However, for target states with $q>0$ the liquid mixture in  Ref.~\cite{jacobs2021self} has zero capacity (see Sec.~\ref{app:jacobscale} of the SI), with the capacity  defined by    $\alpha=\lim_{N\to\infty}p_{\rm max}(N)/N$ and with $p_{\rm max}(N)$  the maximum number of targets that are simultaneously stable; note that this follows conventional definitions of capacities in associative neural networks~\cite{hertz2018introduction}. Instead,  for the liquid Hopfield model the capacity  $\alpha \approx 1$.  This confirms that nonlinearities significantly improve the retrieval capabilities of heterogeneous liquid mixtures and are necessary at nonzero capacity.    Therefore, also the  results in Ref.~\cite{jacobs2021self} are consistent  with  the  trade-off between localisation and retrieval in liquid mixtures (even though it corresponds in our context with a singular  limit with zero capacity).

{\bf Relationship to previous work.}  We further comment on the relation between the current work and previous papers.

The liquid Hopfield model belongs to a family of classical statistical physics models for multi-component mixtures, $N\gg1$, that rely on free energies of the form Eqs.~(\ref{eq:f}-\ref{eq:u}).   While previous work modelled the pairwise affinity matrix as a random matrix, see  e.g., Refs.~\cite{sear2003instabilities, jacobs2013predicting, jacobs2017phase,  carugno2022instabilities, girard2022kinetics, sollich}, here we have established an expression for the pairwise affinity matrix, $J_{ij}$, as a function of a set of prescribed target compositions, $\vec{\xi}^\alpha$. This expression guarantees stability of up to $p_{\rm max}=N-1$ arbitrary targets enriched in any number of components $Q$, and hence solves    explicitly the  inverse problem investigated previously in Refs.~\cite{jacobs2021self, zwicker2022evolved}.

Besides recent work on fluids, the liquid Hopfield model is also related to earlier models for \textit{multifarious}  self-assembly of solid structures~\cite{murugan2015multifarious, sartori2020lessons}.   The main difference is that, in those works, the interactions are specific to reflect the spatial geometry of assemblies, whereas in the current liquid model interactions are non-specific.

An important distinction between the approach presented here and previous work~\cite{jacobs2021self, zwicker2022evolved}, is that  the liquid Hopfield model provides explicit analytical expressions for the affinity matrix $J_{ij}$, while \cite{jacobs2021self, zwicker2022evolved} obtain the affinity matrix by numerically solving the inverse problem.  Having knowledge of an explicit expression for $J_{ij}$, even if it is not the only solution to the inverse problem, is beneficial because it allows for an analytical exploration of the retrieval physics.  Notably, in this study we have identified the spinodal lines that mark the stability boundaries of retrieval states, shedding light on the underlying physical concepts of retrieval, especially highlighting the balance between localization and retrieval.  
Taken together,  the liquid Hopfield model is a toy model for the retrieval of target phases in heterogeneous liquid mixtures, akin to the standard Hopfield model    for  associative memory, with its primary benefit over previous methodologies being the explicit knowledge of the affinity matrix $J_{ij}$.

{\bf Robustness of the liquid Hopfield model.} 
The paper's results on the retrieval capabilities of the liquid Hopfield model and the  trade-off between retrieval and localisation  in liquid mixtures are robust  to  changes in the model definition.   To demonstrate the robustness of the main  results  we consider in the SI the following model variations:

First, we consider a model with  chemical potentials $\mu_i$ drawn from a random distribution, modelling a liquid mixture with variable chemical potentials (see Sec.~\ref{app:chem_pert} of SI). Second, we consider models with    nonlinear terms in the energy density $u$ that are different from the one in Eq.~(\ref{eq:u}).   In particular, in the SI we consider nonlinear terms of higher order, variability in the individual strengths of the cubic repulsion for the different components of the mixture, and  repulsion between different components (see SI Sec. \ref{app:alt-nonlin}).    In all of the above cases, the liquid Hopfield model retains its retrieval capabilities.  
Third, we show analytically that  the stable state of a liquid mixture at low temperature is localised for all affinity matrices $J_{ij}$ that are positive definite   (Sec.~\ref{app:posdef} of SI).   Taken together, these robustness tests show that the specific choices we have made in the definition of the liquid Hopfield model are irrelevant and the retrieval mechanisms we described are generic.  In addition, we have demonstrated that the trade-off between  localization and retrieval is a generic feature of multi-component liquids.

{\bf Perspective.} The results we presented open many natural avenues of future research. Within the specific context of the model, we highlight three. First, it remains open how the key results extend to the canonical ensemble, which may be achieved following a similar analytical approach under a different \textit{ansatz}. Second, the potential role of dissipation, e.g., in stabilizing target phases, remains unaddressed. Third, while we assumed a particular set of interactions, we have not addressed how a mixture can acquire these interactions. Clearly, protein affinities are not random, but have instead evolved to comply with a particular function \cite{sartori2020lessons, zwicker2022evolved}. It remains to be seen whether evolutionary dynamics can parsimoniously converge  to the affinity matrix here proposed, recapitulating the success of neuronal learning rules \cite{hertz2018introduction}.

Besides these theoretical implications, we also foresee experimental implications. Although the liquid Hopfield model constitutes a simple paradigm for cytoplasmic aggregates often referred to as biomolecular condensates, it is yet to be shown that its key findings are in agreement with properties of such cellular structures. Besides this, it is also possible that synthetic experimental constructs of liquid mixtures may be engineered to recapitulate such key findings, as it has recently been attempted for multifarious assembly models using DNA origami \cite{evans2022pattern}.

More generally, this work is the first to establish a direct relationship between the theory of neural computation and that of multi-component liquid mixtures (for solid assemblies, a similar path was followed in \cite{murugan2015multifarious, sartori2020lessons}). Given the vibrant recent advances in both areas, this work suggests that different aspects of neural computation may be applied to liquid mixtures. Examples of these are classical information processing capabilities of feed-forward neural networks, e.g. classification, or dynamic properties such as sequential retrieval \cite{hertz2018introduction}, which have recently also been linked to cubic interactions \cite{herron2023robust}.
\vspace{.1in}

\begin{center}
\textbf{MATERIALS AND METHODS}
\end{center}

{\bf Hypersymmetric targets.} 
We  define the hypersymmetric targets $\vec{\gamma}^{\alpha}$.  These target vectors are the  columns of   Sylvester-Hadamard matrices $\mathcal{H}(k)$ of order $M=2^k$ \cite{hedayat1978hadamard}, and hence we first define  the Sylvester-Hadamard matrices.  These matrices are constructed   through the following iteration,  
\begin{align}
  \mathcal{H}(1) &= \begin{bmatrix}
    1 &  1 \\
    1 & -1
  \end{bmatrix}
  ;
  \mathcal{H}({k}) = \begin{bmatrix}
     \mathcal{H}({{k-1}}) &  \mathcal{H}({{k-1}})\\
     \mathcal{H}({{k-1}}) & -\mathcal{H}({{k-1}})
  \end{bmatrix}
,
\end{align}
which results in the first column having all components equal to one.  Hereafter we do not explicitly indicate the $k$ specifying the order $2^k$ of the matrices.

The columns of these matrices form a Boolean group. In particular, for any two columns $u$ and $v$  with matrix entries $\mathcal{H}_{iu}$ and $\mathcal{H}_{iv}$, there exists a third column $w$ such that
\begin{align}\label{eq:had_group}
\mathcal{H}_{iw}=\mathcal{H}_{iu}\mathcal{H}_{iv}.
\end{align}
We can now define the hypersymmetric targets. For $p=M-1$,  excluding the first column $u=1$, the targets are the columns of Sylvester-Hadamard matrices, i.e., $\gamma_i^{\alpha}=\mathcal{H}_{iu}$. For $p<M-1$, the targets are chosen  randomly among the columns $u>1$ of Sylvester-Hadamard matrices ensuring that for each target state $\vec{\gamma}^{\alpha}$ there exists two target states $\vec{\gamma}^{(\beta_1)}$ and $\vec{\gamma}^{(\beta_2)}$ such that~\eqref{eq:had_group} is satisfied, that is:
\begin{align}\label{eq:target_group}
\gamma^{\alpha}_{i}=\gamma_i^{(\beta_1)}\gamma_i^{(\beta_2)}. 
\end{align} 
In Sec.~\ref{app:symmetric} of the SI we prove that this condition is always fulfilled for $p>M/2$.
\newline

{\bf Conditions for chemical equilibrium.}
The conditions for chemical equilibrium for retrieval states of the form   \eqref{eq:ansatz} are given by
\begin{eqnarray}\label{eq:chemEqQ_MM}
  a_{\ast} &=& n \frac{{\rm exp}\left(w \right) -1}{ {1-q} + {q}{\rm exp}\left(w \right)} ,
\\
\frac{1}{\rho_{\ast}} &=& 1 + {\rm exp}\Big[- \mu_N  +\frac{1}{2} v_3 \rho_{\ast}^2 \left( 1 + a_{\ast}^2 \right)  \label{eq:chemEqRhoQ_MM}\\
& & + q \log\left(1 + a_{\ast} \frac{1-q}{n}\right) + (1 -q) \log\left(1 - a_{\ast}\frac{q}{n}\right) \Big],\nonumber 
\end{eqnarray}

where
$w=\frac{1}{n} \left[v_2 \rho_{\ast} a_{\ast} - \frac{1}{2} v_3 \rho_{\ast}^2 \left(2 a_{\ast} + a_{\ast}^2 \frac{1 - 2q}{n}  \right) \right]$ and $n = \sqrt{q(1 - q)}$.   These conditions hold for arbitrary targets $\vec{\gamma}^\alpha$.   Equations (\ref{eq:chemEqQ_MM}) and (\ref{eq:chemEqRhoQ_MM})  reduce to Eq.~\eqref{eq:crit_point_homogeneous} in the particular case of $a_\ast=0$,  and to Eqs.~\eqref{eq:rho_a_minA} and  \eqref{eq:rho_a_minB} for $q=1/2$. A complete derivation of these conditions can be found in Sec. \ref{app:chemeq} of the SI.
\newline

{\bf Conditions for mechanical equilibrium of homogeneous states.} In  Sec. \ref{app:homo_hess} of the SI we diagonalize the Hessian exactly at the homogeneous state. In particular, the  smallest eigenvalue  is given by
\begin{align}\
\lambda_{\rm min} = \frac{N}{\rho} + v_3 N \rho - v_2 N.
\end{align} 
Setting this value to $0$ gives the condition for mechanical stability of the homogeneous state  (corresponding to the equality of Eq. \ref{eq:crit_temp_homogeneous}). This defines the spinodal line shown in Figs. \ref{fig:phase_diag}, \ref{fig:Jijmod} and \ref{fig:sparse}. We remark that instability happens in a vector space spanned by the target compositions $\vec{\gamma}^\alpha$, which suggests incorrectly, that retrieval becomes stable.
\newline

{\bf Conditions for mechanical equilibrium of retrieval states.} In  Sec. \ref{app:min_eig} of the SI we show that for arbitrary sets of target vectors $\left\{\vec{\gamma}^\beta\right\}_{\beta=1,\ldots,p}$ the minimal eigenvalue of the Hessian evaluated at a retrieval state, $\lambda^{\alpha}_{\rm min}$, is bounded by the following expressions:
\begin{align}
\frac{\lambda^{\alpha}_{\rm min}}{N} &\geq  -v_2 + c_1 -c_2, \label{eq:Lmin_1_MM}\\
\frac{\lambda^{\alpha}_{\rm min}}{N}  &\geq -v_2 + c_1  , \label{eq:Lmin_2_MM}
\end{align}
where
\begin{equation}
c_1 = v_3\rho  \left(1- aq/n\right) + \frac{1}{\rho}\frac{1}{1-aq/n}, \label{eq:c1}
\end{equation}
and 
\begin{align}
c_2 =-\frac{v_3\rho a}{n} + \frac{a}{n \rho}\frac{1}{\left(1+a(1-q)/n\right)\left(1-aq/n\right)}.  \label{eq:c2}
\end{align} 

We use the inequalities above to derive sufficient conditions for retrieval. First, we set $a=a_\ast$ and $\rho=\rho_\ast$, where $a_\ast$ and $\rho_\ast$ are given by Eqs. (\ref{eq:chemEqQ_MM}) and (\ref{eq:chemEqRhoQ_MM}) ensuring chemical stability, and then we impose that the right-hand-side of Eqs.~\eqref{eq:Lmin_1_MM} and \eqref{eq:Lmin_2_MM} are greater or equal than zero. This ensures positivity of the smallest eigenvalue $\lambda^{\alpha}_{\rm min}\ge0$.

For  hypersymmetric targets, the equalities in Eqs.~\eqref{eq:Lmin_1_MM} and \eqref{eq:Lmin_2_MM} are attained. This implies that the conditions described above, evaluated for $q=1/2$ and $n=\sqrt{q(1-q)} = 1/2$ (corresponding with Eqs.~\eqref{eq:stability_ansatz} and \eqref{eq:stability_ansatz2}) are both necessary and sufficient conditions for the stability of  retrieval states with hypersymmetric targets. See SI \ref{app:min_eig} for a proof. In a similar way, in SI \ref{app:pLargeLimit}, we show that the equalities in  Eqs.~\eqref{eq:Lmin_1_MM} and \eqref{eq:Lmin_2_MM} are attained   for generic targets when $p$ is large. When $p$ is not large enough, the condition corresponds to a region where all retrieval states are stable, as shown by the green line in Fig. \ref{fig:Jijmod}A.

\vspace{.1in}

\begin{acknowledgments}
GC is supported by the EPSRC Centre for Doctoral Training in Cross-Disciplinary Approaches to Non-Equilibrium Systems (CANES, EP/L015854/1). RBT acknowledges financial support from the Portuguese Foundation for Science and Technology (FCT) under the contract  2022.12272.BD. PS was partly funded by a grant from laCaixa (LCF/BQ/PI21/11830032).
\end{acknowledgments}

\bibliography{liqhop}

\begin{thebibliography}{49}%
\makeatletter
\providecommand \@ifxundefined [1]{%
 \@ifx{#1\undefined}
}%
\providecommand \@ifnum [1]{%
 \ifnum #1\expandafter \@firstoftwo
 \else \expandafter \@secondoftwo
 \fi
}%
\providecommand \@ifx [1]{%
 \ifx #1\expandafter \@firstoftwo
 \else \expandafter \@secondoftwo
 \fi
}%
\providecommand \natexlab [1]{#1}%
\providecommand \enquote  [1]{``#1''}%
\providecommand \bibnamefont  [1]{#1}%
\providecommand \bibfnamefont [1]{#1}%
\providecommand \citenamefont [1]{#1}%
\providecommand \href@noop [0]{\@secondoftwo}%
\providecommand \href [0]{\begingroup \@sanitize@url \@href}%
\providecommand \@href[1]{\@@startlink{#1}\@@href}%
\providecommand \@@href[1]{\endgroup#1\@@endlink}%
\providecommand \@sanitize@url [0]{\catcode `\\12\catcode `\$12\catcode `\&12\catcode `\#12\catcode `\^12\catcode `\_12\catcode `\%12\relax}%
\providecommand \@@startlink[1]{}%
\providecommand \@@endlink[0]{}%
\providecommand \url  [0]{\begingroup\@sanitize@url \@url }%
\providecommand \@url [1]{\endgroup\@href {#1}{\urlprefix }}%
\providecommand \urlprefix  [0]{URL }%
\providecommand \Eprint [0]{\href }%
\providecommand \doibase [0]{https://doi.org/}%
\providecommand \selectlanguage [0]{\@gobble}%
\providecommand \bibinfo  [0]{\@secondoftwo}%
\providecommand \bibfield  [0]{\@secondoftwo}%
\providecommand \translation [1]{[#1]}%
\providecommand \BibitemOpen [0]{}%
\providecommand \bibitemStop [0]{}%
\providecommand \bibitemNoStop [0]{.\EOS\space}%
\providecommand \EOS [0]{\spacefactor3000\relax}%
\providecommand \BibitemShut  [1]{\csname bibitem#1\endcsname}%
\let\auto@bib@innerbib\@empty
\bibitem [{\citenamefont {Sear}\ and\ \citenamefont {Cuesta}(2003)}]{sear2003instabilities}%
  \BibitemOpen
  \bibfield  {author} {\bibinfo {author} {\bibfnamefont {R.~P.}\ \bibnamefont {Sear}}\ and\ \bibinfo {author} {\bibfnamefont {J.~A.}\ \bibnamefont {Cuesta}},\ }\bibfield  {title} {\bibinfo {title} {Instabilities in complex mixtures with a large number of components},\ }\href@noop {} {\bibfield  {journal} {\bibinfo  {journal} {Physical review letters}\ }\textbf {\bibinfo {volume} {91}},\ \bibinfo {pages} {245701} (\bibinfo {year} {2003})}\BibitemShut {NoStop}%
\bibitem [{\citenamefont {Gavin}\ \emph {et~al.}(2002)\citenamefont {Gavin}, \citenamefont {B{\"o}sche}, \citenamefont {Krause}, \citenamefont {Grandi}, \citenamefont {Marzioch}, \citenamefont {Bauer}, \citenamefont {Schultz}, \citenamefont {Rick}, \citenamefont {Michon}, \citenamefont {Cruciat} \emph {et~al.}}]{gavin2002functional}%
  \BibitemOpen
  \bibfield  {author} {\bibinfo {author} {\bibfnamefont {A.-C.}\ \bibnamefont {Gavin}}, \bibinfo {author} {\bibfnamefont {M.}~\bibnamefont {B{\"o}sche}}, \bibinfo {author} {\bibfnamefont {R.}~\bibnamefont {Krause}}, \bibinfo {author} {\bibfnamefont {P.}~\bibnamefont {Grandi}}, \bibinfo {author} {\bibfnamefont {M.}~\bibnamefont {Marzioch}}, \bibinfo {author} {\bibfnamefont {A.}~\bibnamefont {Bauer}}, \bibinfo {author} {\bibfnamefont {J.}~\bibnamefont {Schultz}}, \bibinfo {author} {\bibfnamefont {J.~M.}\ \bibnamefont {Rick}}, \bibinfo {author} {\bibfnamefont {A.-M.}\ \bibnamefont {Michon}}, \bibinfo {author} {\bibfnamefont {C.-M.}\ \bibnamefont {Cruciat}}, \emph {et~al.},\ }\bibfield  {title} {\bibinfo {title} {Functional organization of the yeast proteome by systematic analysis of protein complexes},\ }\href@noop {} {\bibfield  {journal} {\bibinfo  {journal} {Nature}\ }\textbf {\bibinfo {volume} {415}},\ \bibinfo {pages} {141} (\bibinfo {year} {2002})}\BibitemShut {NoStop}%
\bibitem [{\citenamefont {Marsh}\ and\ \citenamefont {Teichmann}(2015)}]{marsh2015structure}%
  \BibitemOpen
  \bibfield  {author} {\bibinfo {author} {\bibfnamefont {J.~A.}\ \bibnamefont {Marsh}}\ and\ \bibinfo {author} {\bibfnamefont {S.~A.}\ \bibnamefont {Teichmann}},\ }\bibfield  {title} {\bibinfo {title} {Structure, dynamics, assembly, and evolution of protein complexes},\ }\href@noop {} {\bibfield  {journal} {\bibinfo  {journal} {Annual review of biochemistry}\ }\textbf {\bibinfo {volume} {84}},\ \bibinfo {pages} {551} (\bibinfo {year} {2015})}\BibitemShut {NoStop}%
\bibitem [{\citenamefont {Brangwynne}\ \emph {et~al.}(2009)\citenamefont {Brangwynne}, \citenamefont {Eckmann}, \citenamefont {Courson}, \citenamefont {Rybarska}, \citenamefont {Hoege}, \citenamefont {Gharakhani}, \citenamefont {J{\"u}licher},\ and\ \citenamefont {Hyman}}]{brangwynne2009germline}%
  \BibitemOpen
  \bibfield  {author} {\bibinfo {author} {\bibfnamefont {C.~P.}\ \bibnamefont {Brangwynne}}, \bibinfo {author} {\bibfnamefont {C.~R.}\ \bibnamefont {Eckmann}}, \bibinfo {author} {\bibfnamefont {D.~S.}\ \bibnamefont {Courson}}, \bibinfo {author} {\bibfnamefont {A.}~\bibnamefont {Rybarska}}, \bibinfo {author} {\bibfnamefont {C.}~\bibnamefont {Hoege}}, \bibinfo {author} {\bibfnamefont {J.}~\bibnamefont {Gharakhani}}, \bibinfo {author} {\bibfnamefont {F.}~\bibnamefont {J{\"u}licher}},\ and\ \bibinfo {author} {\bibfnamefont {A.~A.}\ \bibnamefont {Hyman}},\ }\bibfield  {title} {\bibinfo {title} {Germline p granules are liquid droplets that localize by controlled dissolution/condensation},\ }\href@noop {} {\bibfield  {journal} {\bibinfo  {journal} {Science}\ }\textbf {\bibinfo {volume} {324}},\ \bibinfo {pages} {1729} (\bibinfo {year} {2009})}\BibitemShut {NoStop}%
\bibitem [{\citenamefont {Hyman}\ \emph {et~al.}(2014)\citenamefont {Hyman}, \citenamefont {Weber},\ and\ \citenamefont {J{\"u}licher}}]{hyman2014liquid}%
  \BibitemOpen
  \bibfield  {author} {\bibinfo {author} {\bibfnamefont {A.~A.}\ \bibnamefont {Hyman}}, \bibinfo {author} {\bibfnamefont {C.~A.}\ \bibnamefont {Weber}},\ and\ \bibinfo {author} {\bibfnamefont {F.}~\bibnamefont {J{\"u}licher}},\ }\bibfield  {title} {\bibinfo {title} {Liquid-liquid phase separation in biology},\ }\href@noop {} {\bibfield  {journal} {\bibinfo  {journal} {Annu. Rev. Cell Dev. Biol}\ }\textbf {\bibinfo {volume} {30}},\ \bibinfo {pages} {39} (\bibinfo {year} {2014})}\BibitemShut {NoStop}%
\bibitem [{\citenamefont {Choi}\ \emph {et~al.}(2020)\citenamefont {Choi}, \citenamefont {Holehouse},\ and\ \citenamefont {Pappu}}]{choi2020physical}%
  \BibitemOpen
  \bibfield  {author} {\bibinfo {author} {\bibfnamefont {J.-M.}\ \bibnamefont {Choi}}, \bibinfo {author} {\bibfnamefont {A.~S.}\ \bibnamefont {Holehouse}},\ and\ \bibinfo {author} {\bibfnamefont {R.~V.}\ \bibnamefont {Pappu}},\ }\bibfield  {title} {\bibinfo {title} {Physical principles underlying the complex biology of intracellular phase transitions},\ }\href@noop {} {\bibfield  {journal} {\bibinfo  {journal} {Annual review of biophysics}\ }\textbf {\bibinfo {volume} {49}},\ \bibinfo {pages} {107} (\bibinfo {year} {2020})}\BibitemShut {NoStop}%
\bibitem [{\citenamefont {Berry}\ \emph {et~al.}(2018)\citenamefont {Berry}, \citenamefont {Brangwynne},\ and\ \citenamefont {Haataja}}]{berry2018physical}%
  \BibitemOpen
  \bibfield  {author} {\bibinfo {author} {\bibfnamefont {J.}~\bibnamefont {Berry}}, \bibinfo {author} {\bibfnamefont {C.~P.}\ \bibnamefont {Brangwynne}},\ and\ \bibinfo {author} {\bibfnamefont {M.}~\bibnamefont {Haataja}},\ }\bibfield  {title} {\bibinfo {title} {Physical principles of intracellular organization via active and passive phase transitions},\ }\href@noop {} {\bibfield  {journal} {\bibinfo  {journal} {Reports on Progress in Physics}\ }\textbf {\bibinfo {volume} {81}},\ \bibinfo {pages} {046601} (\bibinfo {year} {2018})}\BibitemShut {NoStop}%
\bibitem [{\citenamefont {Lingwood}\ and\ \citenamefont {Simons}(2010)}]{lingwood2010lipid}%
  \BibitemOpen
  \bibfield  {author} {\bibinfo {author} {\bibfnamefont {D.}~\bibnamefont {Lingwood}}\ and\ \bibinfo {author} {\bibfnamefont {K.}~\bibnamefont {Simons}},\ }\bibfield  {title} {\bibinfo {title} {Lipid rafts as a membrane-organizing principle},\ }\href@noop {} {\bibfield  {journal} {\bibinfo  {journal} {science}\ }\textbf {\bibinfo {volume} {327}},\ \bibinfo {pages} {46} (\bibinfo {year} {2010})}\BibitemShut {NoStop}%
\bibitem [{\citenamefont {Heberle}\ and\ \citenamefont {Feigenson}(2011)}]{heberle2011phase}%
  \BibitemOpen
  \bibfield  {author} {\bibinfo {author} {\bibfnamefont {F.~A.}\ \bibnamefont {Heberle}}\ and\ \bibinfo {author} {\bibfnamefont {G.~W.}\ \bibnamefont {Feigenson}},\ }\bibfield  {title} {\bibinfo {title} {Phase separation in lipid membranes},\ }\href@noop {} {\bibfield  {journal} {\bibinfo  {journal} {Cold Spring Harbor perspectives in biology}\ }\textbf {\bibinfo {volume} {3}},\ \bibinfo {pages} {a004630} (\bibinfo {year} {2011})}\BibitemShut {NoStop}%
\bibitem [{\citenamefont {Murugan}\ \emph {et~al.}(2015)\citenamefont {Murugan}, \citenamefont {Zeravcic}, \citenamefont {Brenner},\ and\ \citenamefont {Leibler}}]{murugan2015multifarious}%
  \BibitemOpen
  \bibfield  {author} {\bibinfo {author} {\bibfnamefont {A.}~\bibnamefont {Murugan}}, \bibinfo {author} {\bibfnamefont {Z.}~\bibnamefont {Zeravcic}}, \bibinfo {author} {\bibfnamefont {M.~P.}\ \bibnamefont {Brenner}},\ and\ \bibinfo {author} {\bibfnamefont {S.}~\bibnamefont {Leibler}},\ }\bibfield  {title} {\bibinfo {title} {Multifarious assembly mixtures: Systems allowing retrieval of diverse stored structures},\ }\href@noop {} {\bibfield  {journal} {\bibinfo  {journal} {Proceedings of the National Academy of Sciences}\ }\textbf {\bibinfo {volume} {112}},\ \bibinfo {pages} {54} (\bibinfo {year} {2015})}\BibitemShut {NoStop}%
\bibitem [{\citenamefont {Sartori}\ and\ \citenamefont {Leibler}(2020)}]{sartori2020lessons}%
  \BibitemOpen
  \bibfield  {author} {\bibinfo {author} {\bibfnamefont {P.}~\bibnamefont {Sartori}}\ and\ \bibinfo {author} {\bibfnamefont {S.}~\bibnamefont {Leibler}},\ }\bibfield  {title} {\bibinfo {title} {Lessons from equilibrium statistical physics regarding the assembly of protein complexes},\ }\href@noop {} {\bibfield  {journal} {\bibinfo  {journal} {Proceedings of the National Academy of Sciences}\ }\textbf {\bibinfo {volume} {117}},\ \bibinfo {pages} {114} (\bibinfo {year} {2020})}\BibitemShut {NoStop}%
\bibitem [{\citenamefont {Jacobs}\ and\ \citenamefont {Frenkel}(2013)}]{jacobs2013predicting}%
  \BibitemOpen
  \bibfield  {author} {\bibinfo {author} {\bibfnamefont {W.~M.}\ \bibnamefont {Jacobs}}\ and\ \bibinfo {author} {\bibfnamefont {D.}~\bibnamefont {Frenkel}},\ }\bibfield  {title} {\bibinfo {title} {Predicting phase behavior in multicomponent mixtures},\ }\href@noop {} {\bibfield  {journal} {\bibinfo  {journal} {The Journal of chemical physics}\ }\textbf {\bibinfo {volume} {139}},\ \bibinfo {pages} {024108} (\bibinfo {year} {2013})}\BibitemShut {NoStop}%
\bibitem [{\citenamefont {Jacobs}\ and\ \citenamefont {Frenkel}(2017)}]{jacobs2017phase}%
  \BibitemOpen
  \bibfield  {author} {\bibinfo {author} {\bibfnamefont {W.~M.}\ \bibnamefont {Jacobs}}\ and\ \bibinfo {author} {\bibfnamefont {D.}~\bibnamefont {Frenkel}},\ }\bibfield  {title} {\bibinfo {title} {Phase transitions in biological systems with many components},\ }\href@noop {} {\bibfield  {journal} {\bibinfo  {journal} {Biophysical journal}\ }\textbf {\bibinfo {volume} {112}},\ \bibinfo {pages} {683} (\bibinfo {year} {2017})}\BibitemShut {NoStop}%
\bibitem [{\citenamefont {Carugno}\ \emph {et~al.}(2022)\citenamefont {Carugno}, \citenamefont {Neri},\ and\ \citenamefont {Vivo}}]{carugno2022instabilities}%
  \BibitemOpen
  \bibfield  {author} {\bibinfo {author} {\bibfnamefont {G.}~\bibnamefont {Carugno}}, \bibinfo {author} {\bibfnamefont {I.}~\bibnamefont {Neri}},\ and\ \bibinfo {author} {\bibfnamefont {P.}~\bibnamefont {Vivo}},\ }\bibfield  {title} {\bibinfo {title} {Instabilities of complex fluids with partially structured and partially random interactions},\ }\href@noop {} {\bibfield  {journal} {\bibinfo  {journal} {Physical Biology}\ }\textbf {\bibinfo {volume} {19}},\ \bibinfo {pages} {056001} (\bibinfo {year} {2022})}\BibitemShut {NoStop}%
\bibitem [{\citenamefont {Girard}(2022)}]{girard2022kinetics}%
  \BibitemOpen
  \bibfield  {author} {\bibinfo {author} {\bibfnamefont {M.}~\bibnamefont {Girard}},\ }\bibfield  {title} {\bibinfo {title} {On kinetics and extreme values in systems with random interactions},\ }\href@noop {} {\bibfield  {journal} {\bibinfo  {journal} {Physical Biology}\ }\textbf {\bibinfo {volume} {20}},\ \bibinfo {pages} {016006} (\bibinfo {year} {2022})}\BibitemShut {NoStop}%
\bibitem [{\citenamefont {Thewes}\ \emph {et~al.}(2023)\citenamefont {Thewes}, \citenamefont {Kr\"uger},\ and\ \citenamefont {Sollich}}]{sollich}%
  \BibitemOpen
  \bibfield  {author} {\bibinfo {author} {\bibfnamefont {F.~C.}\ \bibnamefont {Thewes}}, \bibinfo {author} {\bibfnamefont {M.}~\bibnamefont {Kr\"uger}},\ and\ \bibinfo {author} {\bibfnamefont {P.}~\bibnamefont {Sollich}},\ }\bibfield  {title} {\bibinfo {title} {Composition dependent instabilities in mixtures with many components},\ }\href {https://doi.org/10.1103/PhysRevLett.131.058401} {\bibfield  {journal} {\bibinfo  {journal} {Phys. Rev. Lett.}\ }\textbf {\bibinfo {volume} {131}},\ \bibinfo {pages} {058401} (\bibinfo {year} {2023})}\BibitemShut {NoStop}%
\bibitem [{\citenamefont {Zwicker}\ and\ \citenamefont {Laan}(2022)}]{zwicker2022evolved}%
  \BibitemOpen
  \bibfield  {author} {\bibinfo {author} {\bibfnamefont {D.}~\bibnamefont {Zwicker}}\ and\ \bibinfo {author} {\bibfnamefont {L.}~\bibnamefont {Laan}},\ }\bibfield  {title} {\bibinfo {title} {Evolved interactions stabilize many coexisting phases in multicomponent liquids},\ }\href@noop {} {\bibfield  {journal} {\bibinfo  {journal} {Proceedings of the National Academy of Sciences}\ }\textbf {\bibinfo {volume} {119}},\ \bibinfo {pages} {e2201250119} (\bibinfo {year} {2022})}\BibitemShut {NoStop}%
\bibitem [{\citenamefont {Jacobs}(2021)}]{jacobs2021self}%
  \BibitemOpen
  \bibfield  {author} {\bibinfo {author} {\bibfnamefont {W.~M.}\ \bibnamefont {Jacobs}},\ }\bibfield  {title} {\bibinfo {title} {Self-assembly of biomolecular condensates with shared components},\ }\href@noop {} {\bibfield  {journal} {\bibinfo  {journal} {Physical review letters}\ }\textbf {\bibinfo {volume} {126}},\ \bibinfo {pages} {258101} (\bibinfo {year} {2021})}\BibitemShut {NoStop}%
\bibitem [{\citenamefont {Hopfield}(1982)}]{hopfield1982neural}%
  \BibitemOpen
  \bibfield  {author} {\bibinfo {author} {\bibfnamefont {J.~J.}\ \bibnamefont {Hopfield}},\ }\bibfield  {title} {\bibinfo {title} {Neural networks and physical systems with emergent collective computational abilities.},\ }\href@noop {} {\bibfield  {journal} {\bibinfo  {journal} {Proceedings of the national academy of sciences}\ }\textbf {\bibinfo {volume} {79}},\ \bibinfo {pages} {2554} (\bibinfo {year} {1982})}\BibitemShut {NoStop}%
\bibitem [{\citenamefont {Callen}(1998)}]{callen1998thermodynamics}%
  \BibitemOpen
  \bibfield  {author} {\bibinfo {author} {\bibfnamefont {H.~B.}\ \bibnamefont {Callen}},\ }\href@noop {} {\bibinfo {title} {Thermodynamics and an introduction to thermostatistics}} (\bibinfo {year} {1998})\BibitemShut {NoStop}%
\bibitem [{\citenamefont {Safran}(2003)}]{samsafransbook}%
  \BibitemOpen
  \bibfield  {author} {\bibinfo {author} {\bibfnamefont {S.}~\bibnamefont {Safran}},\ }\href@noop {} {\emph {\bibinfo {title} {Statistical Thermodynamics Of Surfaces, Interfaces, And Membranes}}}\ (\bibinfo  {publisher} {CRC Press.},\ \bibinfo {year} {2003})\BibitemShut {NoStop}%
\bibitem [{\citenamefont {Doi}(2013)}]{doi2013soft}%
  \BibitemOpen
  \bibfield  {author} {\bibinfo {author} {\bibfnamefont {M.}~\bibnamefont {Doi}},\ }\href@noop {} {\emph {\bibinfo {title} {Soft matter physics}}}\ (\bibinfo  {publisher} {Oxford University Press},\ \bibinfo {year} {2013})\BibitemShut {NoStop}%
\bibitem [{\citenamefont {Mao}\ \emph {et~al.}(2019)\citenamefont {Mao}, \citenamefont {Kuldinow}, \citenamefont {Haataja},\ and\ \citenamefont {Ko{\v{s}}mrlj}}]{mao2019phase}%
  \BibitemOpen
  \bibfield  {author} {\bibinfo {author} {\bibfnamefont {S.}~\bibnamefont {Mao}}, \bibinfo {author} {\bibfnamefont {D.}~\bibnamefont {Kuldinow}}, \bibinfo {author} {\bibfnamefont {M.~P.}\ \bibnamefont {Haataja}},\ and\ \bibinfo {author} {\bibfnamefont {A.}~\bibnamefont {Ko{\v{s}}mrlj}},\ }\bibfield  {title} {\bibinfo {title} {Phase behavior and morphology of multicomponent liquid mixtures},\ }\href@noop {} {\bibfield  {journal} {\bibinfo  {journal} {Soft Matter}\ }\textbf {\bibinfo {volume} {15}},\ \bibinfo {pages} {1297} (\bibinfo {year} {2019})}\BibitemShut {NoStop}%
\bibitem [{\citenamefont {Rubinstein}\ and\ \citenamefont {Colby}(2003)}]{rubinstein2003polymer}%
  \BibitemOpen
  \bibfield  {author} {\bibinfo {author} {\bibfnamefont {M.}~\bibnamefont {Rubinstein}}\ and\ \bibinfo {author} {\bibfnamefont {R.~H.}\ \bibnamefont {Colby}},\ }\href@noop {} {\emph {\bibinfo {title} {Polymer physics}}}\ (\bibinfo  {publisher} {Oxford university press},\ \bibinfo {year} {2003})\BibitemShut {NoStop}%
\bibitem [{\citenamefont {Phillips}(2020)}]{phillips2020molecular}%
  \BibitemOpen
  \bibfield  {author} {\bibinfo {author} {\bibfnamefont {R.}~\bibnamefont {Phillips}},\ }\href@noop {} {\emph {\bibinfo {title} {The molecular switch: Signaling and Allostery}}}\ (\bibinfo  {publisher} {Princeton University Press},\ \bibinfo {year} {2020})\BibitemShut {NoStop}%
\bibitem [{\citenamefont {Luo}\ \emph {et~al.}(2024)\citenamefont {Luo}, \citenamefont {Qiang},\ and\ \citenamefont {Zwicker}}]{luo2024beyond}%
  \BibitemOpen
  \bibfield  {author} {\bibinfo {author} {\bibfnamefont {C.}~\bibnamefont {Luo}}, \bibinfo {author} {\bibfnamefont {Y.}~\bibnamefont {Qiang}},\ and\ \bibinfo {author} {\bibfnamefont {D.}~\bibnamefont {Zwicker}},\ }\bibfield  {title} {\bibinfo {title} {Beyond pairwise: Higher-order physical interactions affect phase separation in multi-component liquids},\ }\href@noop {} {\bibfield  {journal} {\bibinfo  {journal} {arXiv preprint arXiv:2403.06666}\ } (\bibinfo {year} {2024})}\BibitemShut {NoStop}%
\bibitem [{\citenamefont {Hertz}\ \emph {et~al.}(2018)\citenamefont {Hertz}, \citenamefont {Krogh},\ and\ \citenamefont {Palmer}}]{hertz2018introduction}%
  \BibitemOpen
  \bibfield  {author} {\bibinfo {author} {\bibfnamefont {J.}~\bibnamefont {Hertz}}, \bibinfo {author} {\bibfnamefont {A.}~\bibnamefont {Krogh}},\ and\ \bibinfo {author} {\bibfnamefont {R.~G.}\ \bibnamefont {Palmer}},\ }\href@noop {} {\emph {\bibinfo {title} {Introduction to the theory of neural computation}}}\ (\bibinfo  {publisher} {CRC Press},\ \bibinfo {year} {2018})\BibitemShut {NoStop}%
\bibitem [{\citenamefont {Shrinivas}\ and\ \citenamefont {Brenner}(2021)}]{brenner}%
  \BibitemOpen
  \bibfield  {author} {\bibinfo {author} {\bibfnamefont {K.}~\bibnamefont {Shrinivas}}\ and\ \bibinfo {author} {\bibfnamefont {M.~P.}\ \bibnamefont {Brenner}},\ }\bibfield  {title} {\bibinfo {title} {Phase separation in fluids with many interacting components},\ }\href@noop {} {\bibfield  {journal} {\bibinfo  {journal} {Proceedings of the National Academy of Sciences}\ }\textbf {\bibinfo {volume} {118}},\ \bibinfo {pages} {e2108551118} (\bibinfo {year} {2021})}\BibitemShut {NoStop}%
\bibitem [{\citenamefont {Graf}\ and\ \citenamefont {Machta}(2022)}]{machta}%
  \BibitemOpen
  \bibfield  {author} {\bibinfo {author} {\bibfnamefont {I.~R.}\ \bibnamefont {Graf}}\ and\ \bibinfo {author} {\bibfnamefont {B.~B.}\ \bibnamefont {Machta}},\ }\bibfield  {title} {\bibinfo {title} {Thermodynamic stability and critical points in multicomponent mixtures with structured interactions},\ }\href {https://doi.org/10.1103/PhysRevResearch.4.033144} {\bibfield  {journal} {\bibinfo  {journal} {Phys. Rev. Res.}\ }\textbf {\bibinfo {volume} {4}},\ \bibinfo {pages} {033144} (\bibinfo {year} {2022})}\BibitemShut {NoStop}%
\bibitem [{\citenamefont {Kramer}\ and\ \citenamefont {MacKinnon}(1993)}]{kramer1993localization}%
  \BibitemOpen
  \bibfield  {author} {\bibinfo {author} {\bibfnamefont {B.}~\bibnamefont {Kramer}}\ and\ \bibinfo {author} {\bibfnamefont {A.}~\bibnamefont {MacKinnon}},\ }\bibfield  {title} {\bibinfo {title} {Localization: theory and experiment},\ }\href@noop {} {\bibfield  {journal} {\bibinfo  {journal} {Reports on Progress in Physics}\ }\textbf {\bibinfo {volume} {56}},\ \bibinfo {pages} {1469} (\bibinfo {year} {1993})}\BibitemShut {NoStop}%
\bibitem [{\citenamefont {Efetov}(1999)}]{efetov1999supersymmetry}%
  \BibitemOpen
  \bibfield  {author} {\bibinfo {author} {\bibfnamefont {K.}~\bibnamefont {Efetov}},\ }\href@noop {} {\emph {\bibinfo {title} {Supersymmetry in disorder and chaos}}}\ (\bibinfo  {publisher} {Cambridge university press},\ \bibinfo {year} {1999})\BibitemShut {NoStop}%
\bibitem [{\citenamefont {Tsodyks}\ and\ \citenamefont {Feigel'man}(1988)}]{tsodyks1988enhanced}%
  \BibitemOpen
  \bibfield  {author} {\bibinfo {author} {\bibfnamefont {M.~V.}\ \bibnamefont {Tsodyks}}\ and\ \bibinfo {author} {\bibfnamefont {M.~V.}\ \bibnamefont {Feigel'man}},\ }\bibfield  {title} {\bibinfo {title} {The enhanced storage capacity in neural networks with low activity level},\ }\href@noop {} {\bibfield  {journal} {\bibinfo  {journal} {Europhysics Letters}\ }\textbf {\bibinfo {volume} {6}},\ \bibinfo {pages} {101} (\bibinfo {year} {1988})}\BibitemShut {NoStop}%
\bibitem [{\citenamefont {Simons}\ and\ \citenamefont {Toomre}(2000)}]{simons2000lipid}%
  \BibitemOpen
  \bibfield  {author} {\bibinfo {author} {\bibfnamefont {K.}~\bibnamefont {Simons}}\ and\ \bibinfo {author} {\bibfnamefont {D.}~\bibnamefont {Toomre}},\ }\bibfield  {title} {\bibinfo {title} {Lipid rafts and signal transduction},\ }\href@noop {} {\bibfield  {journal} {\bibinfo  {journal} {Nature reviews Molecular cell biology}\ }\textbf {\bibinfo {volume} {1}},\ \bibinfo {pages} {31} (\bibinfo {year} {2000})}\BibitemShut {NoStop}%
\bibitem [{\citenamefont {Chong}\ and\ \citenamefont {Forman-Kay}(2016)}]{chong2016liquid}%
  \BibitemOpen
  \bibfield  {author} {\bibinfo {author} {\bibfnamefont {P.~A.}\ \bibnamefont {Chong}}\ and\ \bibinfo {author} {\bibfnamefont {J.~D.}\ \bibnamefont {Forman-Kay}},\ }\bibfield  {title} {\bibinfo {title} {Liquid--liquid phase separation in cellular signaling systems},\ }\href@noop {} {\bibfield  {journal} {\bibinfo  {journal} {Current opinion in structural biology}\ }\textbf {\bibinfo {volume} {41}},\ \bibinfo {pages} {180} (\bibinfo {year} {2016})}\BibitemShut {NoStop}%
\bibitem [{\citenamefont {Shin}\ and\ \citenamefont {Brangwynne}(2017)}]{shin2017liquid}%
  \BibitemOpen
  \bibfield  {author} {\bibinfo {author} {\bibfnamefont {Y.}~\bibnamefont {Shin}}\ and\ \bibinfo {author} {\bibfnamefont {C.~P.}\ \bibnamefont {Brangwynne}},\ }\bibfield  {title} {\bibinfo {title} {Liquid phase condensation in cell physiology and disease},\ }\href@noop {} {\bibfield  {journal} {\bibinfo  {journal} {Science}\ }\textbf {\bibinfo {volume} {357}},\ \bibinfo {pages} {eaaf4382} (\bibinfo {year} {2017})}\BibitemShut {NoStop}%
\bibitem [{\citenamefont {Banani}\ \emph {et~al.}(2017)\citenamefont {Banani}, \citenamefont {Lee}, \citenamefont {Hyman},\ and\ \citenamefont {Rosen}}]{banani2017biomolecular}%
  \BibitemOpen
  \bibfield  {author} {\bibinfo {author} {\bibfnamefont {S.~F.}\ \bibnamefont {Banani}}, \bibinfo {author} {\bibfnamefont {H.~O.}\ \bibnamefont {Lee}}, \bibinfo {author} {\bibfnamefont {A.~A.}\ \bibnamefont {Hyman}},\ and\ \bibinfo {author} {\bibfnamefont {M.~K.}\ \bibnamefont {Rosen}},\ }\bibfield  {title} {\bibinfo {title} {Biomolecular condensates: organizers of cellular biochemistry},\ }\href@noop {} {\bibfield  {journal} {\bibinfo  {journal} {Nature reviews Molecular cell biology}\ }\textbf {\bibinfo {volume} {18}},\ \bibinfo {pages} {285} (\bibinfo {year} {2017})}\BibitemShut {NoStop}%
\bibitem [{\citenamefont {Murty}\ and\ \citenamefont {Kabadi}(1985)}]{murty1985some}%
  \BibitemOpen
  \bibfield  {author} {\bibinfo {author} {\bibfnamefont {K.~G.}\ \bibnamefont {Murty}}\ and\ \bibinfo {author} {\bibfnamefont {S.~N.}\ \bibnamefont {Kabadi}},\ }\href@noop {} {\emph {\bibinfo {title} {Some NP-complete problems in quadratic and nonlinear programming}}},\ \bibinfo {type} {Tech. Rep.}\ (\bibinfo {year} {1985})\BibitemShut {NoStop}%
\bibitem [{\citenamefont {Pardalos}\ and\ \citenamefont {Schnitger}(1988)}]{pardalos1988checking}%
  \BibitemOpen
  \bibfield  {author} {\bibinfo {author} {\bibfnamefont {P.~M.}\ \bibnamefont {Pardalos}}\ and\ \bibinfo {author} {\bibfnamefont {G.}~\bibnamefont {Schnitger}},\ }\bibfield  {title} {\bibinfo {title} {Checking local optimality in constrained quadratic programming is np-hard},\ }\href@noop {} {\bibfield  {journal} {\bibinfo  {journal} {Operations Research Letters}\ }\textbf {\bibinfo {volume} {7}},\ \bibinfo {pages} {33} (\bibinfo {year} {1988})}\BibitemShut {NoStop}%
\bibitem [{\citenamefont {Pardalos}\ and\ \citenamefont {Vavasis}(1991)}]{pardalos1991quadratic}%
  \BibitemOpen
  \bibfield  {author} {\bibinfo {author} {\bibfnamefont {P.~M.}\ \bibnamefont {Pardalos}}\ and\ \bibinfo {author} {\bibfnamefont {S.~A.}\ \bibnamefont {Vavasis}},\ }\bibfield  {title} {\bibinfo {title} {Quadratic programming with one negative eigenvalue is np-hard},\ }\href@noop {} {\bibfield  {journal} {\bibinfo  {journal} {Journal of Global optimization}\ }\textbf {\bibinfo {volume} {1}},\ \bibinfo {pages} {15} (\bibinfo {year} {1991})}\BibitemShut {NoStop}%
\bibitem [{\citenamefont {M{\'e}zard}\ \emph {et~al.}(1987)\citenamefont {M{\'e}zard}, \citenamefont {Parisi},\ and\ \citenamefont {Virasoro}}]{mezard1987spin}%
  \BibitemOpen
  \bibfield  {author} {\bibinfo {author} {\bibfnamefont {M.}~\bibnamefont {M{\'e}zard}}, \bibinfo {author} {\bibfnamefont {G.}~\bibnamefont {Parisi}},\ and\ \bibinfo {author} {\bibfnamefont {M.~A.}\ \bibnamefont {Virasoro}},\ }\href@noop {} {\emph {\bibinfo {title} {Spin glass theory and beyond: An Introduction to the Replica Method and Its Applications}}},\ Vol.~\bibinfo {volume} {9}\ (\bibinfo  {publisher} {World Scientific Publishing Company},\ \bibinfo {year} {1987})\BibitemShut {NoStop}%
\bibitem [{\citenamefont {Charbonneau}\ \emph {et~al.}(2023)\citenamefont {Charbonneau}, \citenamefont {Marinari}, \citenamefont {Parisi}, \citenamefont {Ricci-tersenghi}, \citenamefont {Sicuro}, \citenamefont {Zamponi},\ and\ \citenamefont {Mezard}}]{charbonneau2023spin}%
  \BibitemOpen
  \bibfield  {author} {\bibinfo {author} {\bibfnamefont {P.}~\bibnamefont {Charbonneau}}, \bibinfo {author} {\bibfnamefont {E.}~\bibnamefont {Marinari}}, \bibinfo {author} {\bibfnamefont {G.}~\bibnamefont {Parisi}}, \bibinfo {author} {\bibfnamefont {F.}~\bibnamefont {Ricci-tersenghi}}, \bibinfo {author} {\bibfnamefont {G.}~\bibnamefont {Sicuro}}, \bibinfo {author} {\bibfnamefont {F.}~\bibnamefont {Zamponi}},\ and\ \bibinfo {author} {\bibfnamefont {M.}~\bibnamefont {Mezard}},\ }\href@noop {} {\emph {\bibinfo {title} {Spin Glass Theory and Far Beyond: Replica Symmetry Breaking after 40 Years}}}\ (\bibinfo  {publisher} {World Scientific},\ \bibinfo {year} {2023})\BibitemShut {NoStop}%
\bibitem [{\citenamefont {Hopfield}(1984)}]{hopfield1984neurons}%
  \BibitemOpen
  \bibfield  {author} {\bibinfo {author} {\bibfnamefont {J.~J.}\ \bibnamefont {Hopfield}},\ }\bibfield  {title} {\bibinfo {title} {Neurons with graded response have collective computational properties like those of two-state neurons.},\ }\href@noop {} {\bibfield  {journal} {\bibinfo  {journal} {Proceedings of the national academy of sciences}\ }\textbf {\bibinfo {volume} {81}},\ \bibinfo {pages} {3088} (\bibinfo {year} {1984})}\BibitemShut {NoStop}%
\bibitem [{\citenamefont {Boll{\'e}}\ \emph {et~al.}(2003)\citenamefont {Boll{\'e}}, \citenamefont {Nieuwenhuizen}, \citenamefont {Castillo},\ and\ \citenamefont {Verbeiren}}]{bolle2003spherical}%
  \BibitemOpen
  \bibfield  {author} {\bibinfo {author} {\bibfnamefont {D.}~\bibnamefont {Boll{\'e}}}, \bibinfo {author} {\bibfnamefont {T.~M.}\ \bibnamefont {Nieuwenhuizen}}, \bibinfo {author} {\bibfnamefont {I.~P.}\ \bibnamefont {Castillo}},\ and\ \bibinfo {author} {\bibfnamefont {T.}~\bibnamefont {Verbeiren}},\ }\bibfield  {title} {\bibinfo {title} {A spherical hopfield model},\ }\href@noop {} {\bibfield  {journal} {\bibinfo  {journal} {Journal of Physics A: Mathematical and General}\ }\textbf {\bibinfo {volume} {36}},\ \bibinfo {pages} {10269} (\bibinfo {year} {2003})}\BibitemShut {NoStop}%
\bibitem [{\citenamefont {McGraw}\ and\ \citenamefont {Menzinger}(2003{\natexlab{a}})}]{Mcgraw1}%
  \BibitemOpen
  \bibfield  {author} {\bibinfo {author} {\bibfnamefont {P.~N.}\ \bibnamefont {McGraw}}\ and\ \bibinfo {author} {\bibfnamefont {M.}~\bibnamefont {Menzinger}},\ }\bibfield  {title} {\bibinfo {title} {Bistable gradient networks. i. attractors and pattern retrieval at low loading in the thermodynamic limit},\ }\href {https://doi.org/10.1103/PhysRevE.67.016118} {\bibfield  {journal} {\bibinfo  {journal} {Phys. Rev. E}\ }\textbf {\bibinfo {volume} {67}},\ \bibinfo {pages} {016118} (\bibinfo {year} {2003}{\natexlab{a}})}\BibitemShut {NoStop}%
\bibitem [{\citenamefont {McGraw}\ and\ \citenamefont {Menzinger}(2003{\natexlab{b}})}]{Mcgraw2}%
  \BibitemOpen
  \bibfield  {author} {\bibinfo {author} {\bibfnamefont {P.~N.}\ \bibnamefont {McGraw}}\ and\ \bibinfo {author} {\bibfnamefont {M.}~\bibnamefont {Menzinger}},\ }\bibfield  {title} {\bibinfo {title} {Bistable gradient networks. ii. storage capacity and behavior near saturation},\ }\href {https://doi.org/10.1103/PhysRevE.67.016119} {\bibfield  {journal} {\bibinfo  {journal} {Phys. Rev. E}\ }\textbf {\bibinfo {volume} {67}},\ \bibinfo {pages} {016119} (\bibinfo {year} {2003}{\natexlab{b}})}\BibitemShut {NoStop}%
\bibitem [{\citenamefont {Castellani}\ and\ \citenamefont {Cavagna}(2005)}]{castellani2005spin}%
  \BibitemOpen
  \bibfield  {author} {\bibinfo {author} {\bibfnamefont {T.}~\bibnamefont {Castellani}}\ and\ \bibinfo {author} {\bibfnamefont {A.}~\bibnamefont {Cavagna}},\ }\bibfield  {title} {\bibinfo {title} {Spin-glass theory for pedestrians},\ }\href@noop {} {\bibfield  {journal} {\bibinfo  {journal} {Journal of Statistical Mechanics: Theory and Experiment}\ }\textbf {\bibinfo {volume} {2005}},\ \bibinfo {pages} {P05012} (\bibinfo {year} {2005})}\BibitemShut {NoStop}%
\bibitem [{\citenamefont {Evans}\ \emph {et~al.}(2022)\citenamefont {Evans}, \citenamefont {O'Brien}, \citenamefont {Winfree},\ and\ \citenamefont {Murugan}}]{evans2022pattern}%
  \BibitemOpen
  \bibfield  {author} {\bibinfo {author} {\bibfnamefont {C.~G.}\ \bibnamefont {Evans}}, \bibinfo {author} {\bibfnamefont {J.}~\bibnamefont {O'Brien}}, \bibinfo {author} {\bibfnamefont {E.}~\bibnamefont {Winfree}},\ and\ \bibinfo {author} {\bibfnamefont {A.}~\bibnamefont {Murugan}},\ }\bibfield  {title} {\bibinfo {title} {Pattern recognition in the nucleation kinetics of non-equilibrium self-assembly},\ }\href@noop {} {\bibfield  {journal} {\bibinfo  {journal} {arXiv preprint arXiv:2207.06399}\ } (\bibinfo {year} {2022})}\BibitemShut {NoStop}%
\bibitem [{\citenamefont {Herron}\ \emph {et~al.}(2023)\citenamefont {Herron}, \citenamefont {Sartori},\ and\ \citenamefont {Xue}}]{herron2023robust}%
  \BibitemOpen
  \bibfield  {author} {\bibinfo {author} {\bibfnamefont {L.}~\bibnamefont {Herron}}, \bibinfo {author} {\bibfnamefont {P.}~\bibnamefont {Sartori}},\ and\ \bibinfo {author} {\bibfnamefont {B.}~\bibnamefont {Xue}},\ }\bibfield  {title} {\bibinfo {title} {Robust retrieval of dynamic sequences through interaction modulation},\ }\href@noop {} {\bibfield  {journal} {\bibinfo  {journal} {PRX Life}\ }\textbf {\bibinfo {volume} {1}},\ \bibinfo {pages} {023012} (\bibinfo {year} {2023})}\BibitemShut {NoStop}%
\bibitem [{\citenamefont {Hedayat}\ and\ \citenamefont {Wallis}(1978)}]{hedayat1978hadamard}%
  \BibitemOpen
  \bibfield  {author} {\bibinfo {author} {\bibfnamefont {A.}~\bibnamefont {Hedayat}}\ and\ \bibinfo {author} {\bibfnamefont {W.~D.}\ \bibnamefont {Wallis}},\ }\bibfield  {title} {\bibinfo {title} {Hadamard matrices and their applications},\ }\href@noop {} {\bibfield  {journal} {\bibinfo  {journal} {The Annals of Statistics}\ ,\ \bibinfo {pages} {1184}} (\bibinfo {year} {1978})}\BibitemShut {NoStop}%
\end{thebibliography}%
\clearpage

\onecolumngrid

\begin{center}
    \Large{{\bf Supplementary Information}}
\end{center}
\section{Hypersymmetric targets for  $p>M/2$}
\label{app:symmetric}

For $p>M/2$, the  condition (\ref{eq:target_group}) is  satisfied for all combinations of $p$ targets.    Indeed, this can be proven through contradiction.   Assume that there exists one target state, say $\vec{\gamma}^{\alpha}$, for which there exist no two other target states $\vec{\gamma}^{(\beta_1)}$ and $\vec{\gamma}^{(\beta_2)}$  such that (\ref{eq:target_group}) holds.   In this case, we can construct the $p-1$ vectors with entries  $\zeta^{\beta}_i = \gamma^{\alpha}_{i}\gamma^{\beta}_{i}$ where $\beta\neq \alpha$, which by assumption  are not target states.   However, since the columns of the Hadamard matrix form a group under the element wise product,  the $\vec{\zeta}^{\beta}$, with $\beta\neq\alpha$, are $p-1$ distinct columns of the Hadamard matrix.    Hence,  within our assumptions, the $p-1$ vectors $\vec{\zeta}^{\beta}$ with $\beta \neq \alpha$, the $p-1$ target states  $\vec{\gamma}^{\beta}$  with $\beta \neq \alpha$, the target state  $\vec{\gamma}^{\alpha}$, and the all-ones vector,  should all be   distinct  columns of the Hadamard matrix.   However, this is not possible as we count $2p>M$ vectors  and the Hadamard matrix has by construction only $M$ columns.

\section{Global stability lines}
\label{app:binodal}

\begin{figure}[h]
\centerline{\includegraphics[width = 10cm]{ 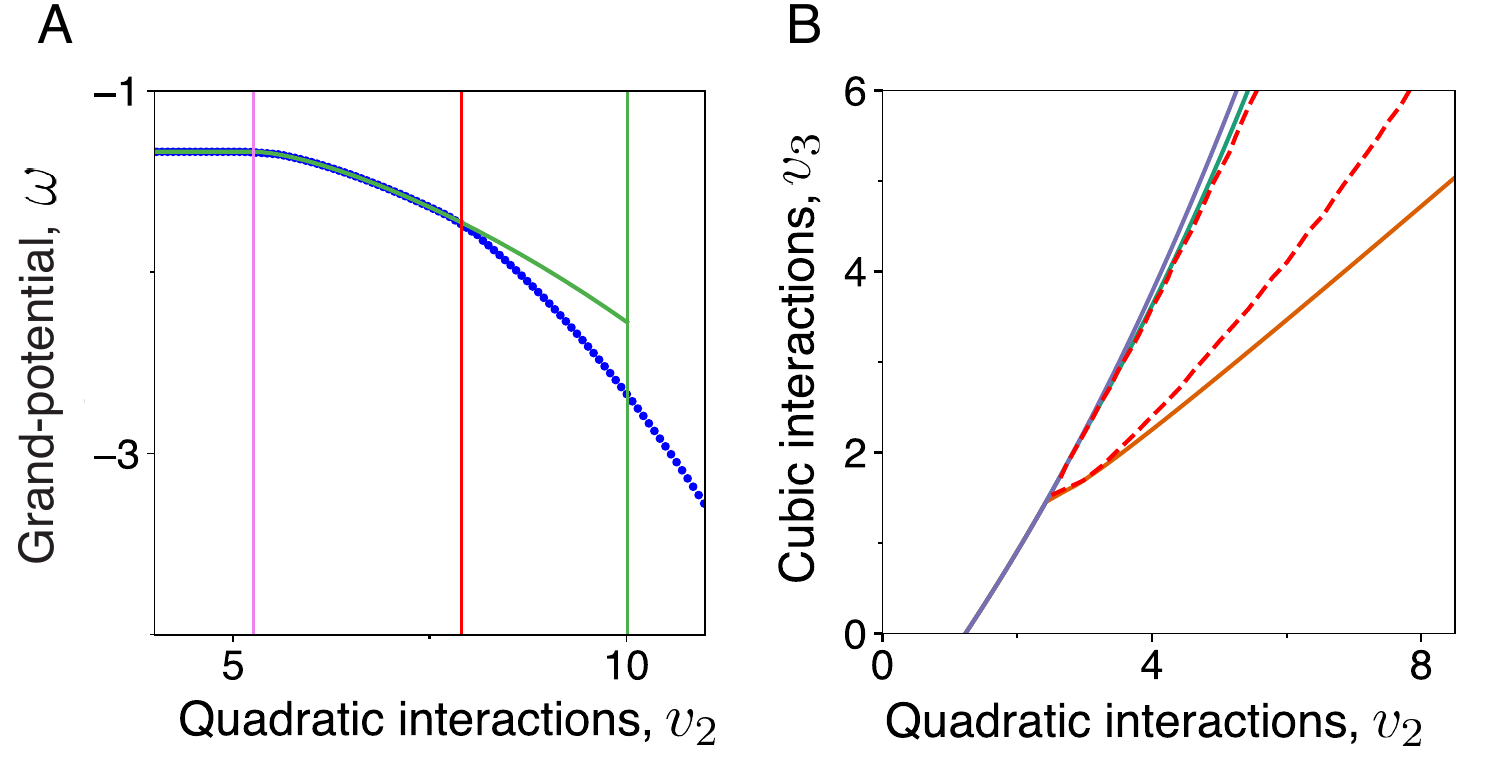}}
\caption{\label{fig:Grand-potential_minima}\textit{Global stability lines of retrieval phase.} Panel A:  Grand-potential $\omega$ of the locally stable retrieval state (green solid line) and the globally stable state (blue markers) as a function of  $v_2$.   The vertical lines denote, from left to right, the spinodal  when the homogeneous region becomes unstable (pink),  the global stability line when retrieval becomes globally unstable (red), and the spinodal when the retrieval state becomes unstable (green).  Cubic interactions are set to $v_3 = 3$.  Panel B: Stability diagram showing both the spinodals (solid lines) and the global stability lines (dashed lines).   Parameters are set to $\mu_N =1.5 $, $N = 32$ and $p = 6$. }
\end{figure}

In the main text, we have focused on local stability of retrieval states.  Therefore, we have  calculated the spinodal lines of the retrieval phase. Here, we discuss global stability lines, which delimit the region where the retrieval states are not only local minima of the grand-potential functional, but are its global minima. In Fig.~\ref{fig:Grand-potential_minima}A we compare the  grand-potential functional evaluated at the retrieval state (green line) with the grand-potential, i.e. the global-minimum of the grand-potential functional identified by numerical minimization (blue markers). The figure shows a region for which the retrieval state is globally stable, and a region for which the retrieval state is locally stable but not globally stable. The global stability line delimits the former, and the spinodal delimits the latter region.  Figure~\ref{fig:Grand-potential_minima}  shows  a stability diagram displaying both spinodals and global stability lines. As it is readily seen, the global stability lines show qualitatively similar trends as the spinodals. 
\section{Conditions of chemical equilibrium for homogeneous and retrieval states}
\label{app:chemeq}
Here, we derive the conditions under which the \textit{ansatz} in Eq.~\eqref{eq:ansatz} is in chemical equilibrium. We derive these conditions for affinity matrices of the form~\eqref{eq:Jij}, and these conditions reduce to~\eqref{eq:crit_point_homogeneous} for $a_\ast=0$ and to  Eqs.~\eqref{eq:rho_a_minA} and  \eqref{eq:rho_a_minB} for $q=1/2$.

The condition for chemical equilibrium in Eq.~\eqref{eq:locmin} can be written explicitly as
\begin{equation}\label{eq:b1}
-v_2 \sum^N_{j=1} J_{ij}\rho_j + \frac{v_3}{2}\rho^2_i N^2 + \log \rho_i = \log (1-\rho) + \mu^{\rm res},
\end{equation} 
which yields $N$ coupled equations. Substituting the retrieval \textit{ansatz} given by Eq.~\eqref{eq:ansatz} into \eqref{eq:b1} gives 
\begin{align}\label{eq:chemeqapp}
-v_2 \sum^N_{j=1} J_{ij}\frac{\rho_\ast}{N} (1+a_\ast\gamma^{\alpha}_j)  + \frac{v_3}{2}\rho^2_\ast  (1+a_\ast\gamma^{\alpha}_i)^2  + \log \frac{\rho_\ast(1+a_\ast\gamma^{\alpha}_i)}{N}=\log (1-\rho_\ast) + \mu^{\rm res} .
\end{align} 
We now use two properties of the affinity matrix $J_{ij}$ of Eq.~\eqref{eq:Jij}: first, the all-ones vector belongs to the kernel of $J_{ij}$; and second, the target vectors, $\vec{\gamma}^{\alpha}$, are eigenvectors of $J_{ij}$ with eigenvalue $N$. From this, we can simplify Eq.~\eqref{eq:chemeqapp} into
\begin{align}\label{eq:gen_a_rho_ast}
-v_2 {\rho_\ast} a _\ast\gamma^{\alpha}_i  + \frac{v_3}{2}\rho^2 _\ast  (1+a _\ast\gamma^{\alpha}_i)^2+ \log \frac{\rho_\ast(1+a_\ast\gamma^{\alpha}_i)}{N} = \log (1-\rho_\ast) + \mu^{\rm res}.
\end{align}
Note that because the targets are binary, this reduces the $N$ stability equations to two equations:  one for  $\gamma^{\alpha}_i = (1-q)/n$ and the other one for $\gamma^{\alpha}_i = -q/n$. Solving these towards $a$ and $\rho$ we get the chemical equilibrium conditions for target states with arbitrary $q$ (Eq.~\eqref{eq:chemEqQ_MM} and \eqref{eq:chemEqRhoQ_MM}) .

\section{Exact diagonalization of the Hessian in the homogeneous state}
\label{app:homo_hess}
We diagonalize  the Hessian in a homogeneous state $\rho_i=\rho/N$, which takes the form
\begin{align}\label{eq:hess_hom}
&H_{ij}= -  v_2 J_{ij} +  \frac{1}{1-\rho} + \delta_{ij} \left( \frac{N}{\rho} + v_3 N \rho \right) ,
\end{align}
where $J_{ij}$ is given by Eq.~\eqref{eq:Jij}.

The spectrum of this matrix  consists of three eigenvalues, which we denote by $\lambda_{\rm ret}$, $\lambda_{\rm cond}$ and $\lambda_{\perp}$. The retrieval eigenvalue is
\begin{align}
\lambda_{\rm ret} = \frac{N}{\rho} + v_3 N \rho - v_2 N.
\end{align} 
Its eigenspace is a $p$-dimensional vector space spanned by the $p$ targets  $\vec{\gamma}^{\alpha}$ to be retrieved. The  condensation eigenvalue is
\begin{align}
\lambda_{\rm cond}= \frac{N}{\rho} + v_3 N \rho + \frac{N}{1-\rho},
\end{align} 
and its corresponding  eigenvector is the all-ones vector. The remaining eigenvalue is 
\begin{equation}
\lambda_{\perp}=  \frac{N}{\rho} + v_3 N \rho,
\end{equation}
and has a $N-p-1$ dimensional eigenspace that is orthogonal to the targets and the all-ones vector. 

Since only  $\lambda_{\rm ret}$ can be negative, the homogeneous state destabilises in a direction that is a linear combination of the target vectors. The condition in Eq.~\eqref{eq:crit_temp_homogeneous} for mechanical equilibrium is obtained by setting $\rho=\rho_\ast$ according to Eq.~\eqref{eq:crit_point_homogeneous} in the equation $\lambda_{\rm ret}=0$.

\section{Lower bounds for eigenvalues of the Hessian in the retrieval state}
\label{app:min_eig}

Unlike in the homogeneous case, discussed in Sec.~\ref{app:homo_hess}, we do not determine the full spectrum of the hessian in retrieval states. Nevertheless, we  derive a lower bound for the  eigenvalues of the Hessian which provide sufficient conditions for retrieval.  Note that the lower bound applies to a generic set of $p$ target vectors $\vec{\gamma}^\alpha$. Furthermore, we also show that for hypersymmetric targets, as defined in \ref{app:symmetric}, these conditions are in fact also necessary, which proves Eqs.~(\ref{eq:stability_ansatz})  and (\ref{eq:stability_ansatz2}).

Substituting the retrieval ansatz (\ref{eq:ansatz}) into the Hessian, $H_{ij}=\partial^2 f/\partial\rho_i\partial\rho_j$, we get the matrix 
\begin{align}
{H^{\alpha}_{ij} = -v_2 J_{ij} + \frac{1}{1-\rho}}  + N \delta_{i,j}\left[v_3\rho (1+a\gamma^{\alpha}_i) + \frac{1}{\rho}\frac{1}{1+a\gamma^{\alpha}_i}\right].
\end{align} 
Substituting $\gamma^{\alpha}_i=(\xi^{\alpha}_i-q)/n$, and using that $\xi^{\alpha}_i\in \left\{0,1\right\}$, we get 
\begin{equation}
H^\alpha_{ij} = -v_2 J_{ij} + \frac{1}{1-\rho} + N\delta_{i,j}\left(c_1-c_2\xi^{\alpha}_i\right) ,
\end{equation}
where 
\begin{equation}
c_1 = v_3\rho  \left(1- aq/n\right) + \frac{1}{\rho}\frac{1}{1-aq/n}, \label{eq:c1}
\end{equation}
and 
\begin{align}
c_2 =-\frac{v_3\rho a}{n} + \frac{a}{n \rho}\frac{1}{\left(1+a(1-q)/n\right)\left(1-aq/n\right)}.  \label{eq:c2}
\end{align} 
Since the matrix $H^\alpha_{ij}$ is symmetric, its minimal eigenvalue $\lambda^{\alpha}_{\rm min}$ can be obtained by solving the constrained minimization problem 
\begin{equation}
\lambda^{\alpha}_{\rm min} = \underset{\|\vec{x}\| = 1}{\min}\left( \sum^N_{i=1}\sum^N_{j=1}x_iH^\alpha_{ij}x_j\right),  \label{eq:variational}
\end{equation}
where 
\begin{eqnarray}\label{eq:quad}
{\sum^N_{i=1}\sum^N_{j=1}x_iH^\alpha_{ij}x_j} =  -v_2\sum^N_{i=1}\sum^N_{j=1}x_iJ_{ij}x_j + \frac{1}{1-\rho}\left(\sum^N_{i=1}x_i\right)^2  + Nc_1 \sum^N_{i=1}x^2_i - Nc_2 \sum^N_{i=1}x^2_i\xi^{\alpha}_i .\label{eq:quadratic}
\end{eqnarray}

Next, we  bound $\lambda^{\alpha}_{\rm min}$ from below by bounding the four terms in Eq.~(\ref{eq:quadratic}):
\begin{itemize}
\item {\it  interaction} term: since the affinity matrix $J_{ij}$ in Eq.~\eqref{eq:Jij} has two eigenvalues, namely $0$ and $N$, it holds that  
\begin{equation}
 -v_2\sum^N_{i=1}\sum^N_{j=1}J_{ij}x_ix_j \geq  -v_2 N\sum^N_{i=1}x^2_i  = -v_2N.
 \end{equation}

\item  {\it  entropic} term: this term is non-negative, and so
  \begin{equation}
  \frac{1}{1-\rho}\left(\sum^N_{i=1}x_i\right)^2 \geq 0.
   \end{equation}  
\item  {\it  $c_1$}-term: as the norm of $\left\{x_i\right\}$ equals one,  this term is constant,   
\begin{equation}
Nc_1 \sum^N_{i=1}x^2_i =Nc_1.
\end{equation}

\item {\it  $c_2$}-term: depending on the sign of $c_2$ we get different bounds. If $c_2>0$, then 
\begin{equation}
- Nc_2 \sum^N_{i=1}x^2_i\xi^{\alpha}_i \geq - Nc_2.
\end{equation}
On the other hand, if $c_2<0$, then 
\begin{equation}
- Nc_2 \sum^N_{i=1}x^2_i\xi^{\alpha}_i \geq 0.
\end{equation}
\end{itemize}
Adding the lower bounds for the individual terms, we get  for $c_2>0$ that 
\begin{equation}
\frac{\lambda^{\alpha}_{\rm min}}{N} \geq  -v_2 + c_1 -c_2, \label{eq:Lmin_1}
\end{equation}
and for $c_2<0$, 
\begin{equation}
\frac{\lambda^{\alpha}_{\rm min}}{N}  \geq -v_2 + c_1  . \label{eq:Lmin_2}
\end{equation}

The inequalities above can be used to derive sufficient conditions for retrieval, that is conditions on the parameters $\{v_2, v_3, \mu, q\}$ that, when satisfied, ensure stability of the retrieval states (but there may be other regions of parameters where retrieval is also stable). To do so, we set $a=a_\ast$ and $\rho=\rho_\ast$, where $a_\ast$ and $\rho_\ast$ solve Eq.~\eqref{eq:gen_a_rho_ast} ensuring chemical stability, and then we impose that the right-hand-side (RHS) of Eqs.~\eqref{eq:Lmin_1} and \eqref{eq:Lmin_2} are greater or equal than zero. Since positivity of the RHS ensures $\lambda^{\alpha}_{\rm min}\ge0$ we find the following sufficient conditions for stability of retrieval states, 
\begin{align}
-v_2 + c_1 -c_2&\ge0 \label{eq:1Q}
\end{align}
for $c_2>0$, and 
\begin{align}
-v_2 + c_1&\ge0 \label{eq:2Q}
\end{align}
for $c_2<0$.   Note that if $c_2>0$, then (\ref{eq:1Q}) implies (\ref{eq:2Q}), and if $c_2<0$, then (\ref{eq:2Q}) implies (\ref{eq:1Q}).   Hence, we can state mechanical stability conditions more simply as  the requirement that both (\ref{eq:1Q}) and (\ref{eq:2Q})  are satisfied, without reference to $c_2$.
Remarkably, the sufficient conditions (\ref{eq:1Q}) and (\ref{eq:2Q})  for retrieval do not depend on $p$.   Eqs.~\eqref{eq:1Q} and \eqref{eq:2Q}  were used in Fig.~\ref{fig:Jijmod}A to outline the boundary of the retrieval phase with arbitrary sparsity $q$.

For  hypersymmetric targets, as defined in Sec.~\ref{app:symmetric}, the inequalities in Eqs.~\eqref{eq:Lmin_1} and \eqref{eq:Lmin_2} saturate. This implies that the conditions  Eqs.~\eqref{eq:1Q} and \eqref{eq:2Q} evaluated for $q=1/2$ and $n=\sqrt{q(1-q)} = 1/2$ (corresponding with Eqs.~\eqref{eq:stability_ansatz} and \eqref{eq:stability_ansatz2} are both necessary and sufficient conditions for  the stability of  retrieval states with hypersymmetric targets. To show that the equality in  Eq.~\eqref{eq:Lmin_1} is attained when the target vectors are hypersymmetric,    we evaluate the quadratic function in Eq.~\eqref{eq:quad} at
\begin{align}
\vec{x} =\frac{1}{\sqrt{2N}}\left( \vec{\gamma}^{(\beta_1)} + \vec{\gamma}^{(\beta_2)}\right) . \label{eq:xiGammaBeta1}
\end{align}
Using  the property Eq.~\eqref{eq:target_group} that holds for hypersymmetric target vectors, we find that this choice of $\vec{x}$ saturates Eq.~\eqref{eq:Lmin_1}. For $c_2<0$ we obtain saturation using the vector
\begin{align}
\vec{x} =\frac{1}{\sqrt{2N}}\left( \vec{\gamma}^{(\beta_1)} - \vec{\gamma}^{(\beta_2)} \right) .\label{eq:xiGammaBeta2}
\end{align}

\section{Collapse of the retrieval lines in the limit of large $p$ and $q=1/2$}
\label{app:pLargeLimit}
Asides from the symmetric case, the inequalities in Eqs.~\eqref{eq:Lmin_1} and \eqref{eq:Lmin_2} saturate in the limit of a large number of targets $p$. We observed this numerically for all values of $q$, and below we provide an analytical derivation for the case $q=1/2$, which explains why in Fig.~\ref{fig:Jijmod}A the spinodal lines are determined by the equalities in Eq.~(\ref{eq:stability_ansatz}) for large $p$.

To prove  saturation at large values of $p$, we note that the equalities in (\ref{eq:1Q}) and (\ref{eq:2Q}) are attained for the vectors (\ref{eq:xiGammaBeta1}) and (\ref{eq:xiGammaBeta2}) because (i) $\vec{x}$ is an eigenvector of $J_{ij}$; (ii) $\vec{x}$ is a linear combination of the  two vectors $\vec{\gamma}^{(\beta_1)}$  and $\vec{\gamma}^{(\beta_2)}$  for which it holds that $\gamma^{(\beta_1)}_i\gamma^{(\beta_2)}_i = \gamma^{\alpha}_i$ for all $i$.       

Here, we follow a similar approach.  We evaluate the quadratic form in Eq.~\eqref{eq:quad} at the vector
\begin{equation}
\vec{x}^{(+,\alpha)} =\frac{1}{\sqrt{2N}}\left( \vec{\zeta}^{(\beta_1)} + \vec{\zeta}^{(\beta_2)} \right) \label{eq:x2}
\end{equation}
where $\vec{\zeta}^{(\beta_1)}$  and $\vec{\zeta}^{(\beta_2)}$ are two vectors, not necessarily target states, with entries $\zeta^{(\beta_1)}_i, \zeta^{(\beta_2)}_i\in \left\{-1,1\right\}$, for which $\sum_i\zeta^{(\beta_1)}_i = \sum_i\zeta^{(\beta_2)}_i = 0$, and with $\zeta^{(\beta_1)}_i \zeta^{(\beta_2)}_i = \gamma^{\alpha}_i$.       Such two states can be constructed as follows.   First, we select a state $\vec{\zeta}^{(\beta_1)}$ with binary entries that is orthogonal to $
\vec{\gamma}^{\alpha}$ and to the all ones vector.   Second,  we define the vector with entries $\zeta^{(\beta_2)}_i = \zeta^{(\beta_1)}_i\gamma^{\alpha}_i$.   
Since 
 $p=N-1$,  it holds that the affinity matrix $J_{ij} = \delta_{i,j}- 1$,  and therefore $\vec{x}^{(+,\alpha)}$ is an eigenvector of $J_{ij}$.  Consequently, the inequality (\ref{eq:1Q}) is saturated for when setting $\vec{x}=\vec{x}^{(+,\alpha)}$ in the quadratic form (\ref{eq:quadratic}).  
 Analogously, we can consider a vector  $\vec{x}^{(-,\alpha)}$ to obtain the second spinodal line (\ref{eq:2Q}).   

\section{Equilibrium states at zero temperature}
\label{app:zeroT} 
\subsection{Hebbian interaction matrices}
We now use a probabilistic argument to show that for $p>\log(N)/\log(2)$ the minima of the functional, $\omega$ (Eq.~\eqref{eq:omega}) in the limit of zero temperature, are the corners of the simplex of physical states. 
First, note that in the limit of $v_3=0$, $\mu^{\rm res}=0$, and $v_2\gg 1$ the functional takes the quadratic form
\begin{align} \label{eq:QO}
\omega(\vec{\rho},0)=- \frac{v_2}{2} \sum_{i,j=1}^N J_{ij} \rho_i \rho_j. 
\end{align}
Minimising a quadratic function over a simplex is an NP-hard problem called the  quadratic concave optimization problem, see Refs.~\cite{murty1985some, pardalos1988checking,  pardalos1991quadratic}. However, for the specific choice of $J_{ij}$ in Eq.~\eqref{eq:Jij} it holds that
\begin{align}
\omega(\vec{\rho},0)=- \frac{v_2}{2} \rho^2 \sum_{\alpha=1}^{p} m_{\alpha}^2,  \label{eq:QO2} 
\end{align}
where $m_{\alpha} =  \sum^N_{i=1} \gamma^{\alpha}_i \rho_i$.
The minima must have $\rho=1$ and $m_{\alpha}=\pm1$ for all values of $\alpha$. Hence, the minima are obtained as solutions to equations of the form $m_\alpha = \sigma_\alpha$, with $\alpha\in \left\{1,2,\dots,p\right\}$, and $\sigma_\alpha \in \left\{-1,1\right\}$. The values of $\sigma_\alpha$ determine the corner on which we focus. We discuss the case of $\sigma_\alpha=1$ for all $\alpha$, but the other cases can be treated equivalently. The effect of each equation is to set $\rho^{\alpha}_i=0$ whenever $\gamma^{\alpha}_i\neq 1$.  In this way, given a set of $p$ targets,  a noncorner solution exists whenever there exist two or more components, say $i$ and $j$, for which it holds that $\gamma^{\alpha}_i = \gamma^{\alpha}_j=1$ for all $\alpha\in \left\{1,2,\ldots,p\right\}$. That is, whenever two or more components are shared by all targets.
If the $p$ targets are randomly selected, the probability that at least two components are shared in all targets is 
\begin{equation}
p_{++} = \left(\begin{array}{c}N \\ 2\end{array}\right)\frac{1}{2^{2p}}
\end{equation} 
where the first term counts the number of ways of selecting two components out of the N, and the second corresponds to the probability that both components are present in all $p$ targets. We are interested in the behavior of $p_{++}$ when $p,N\rightarrow\infty$. In this limit, we can write $p_{++}\approx \exp(- \zeta)$ with $ \zeta =$ $- 2\log(N)+2p\log(2)$. Calculating when this probability approaches zero, determines when the minima of the functional are exclusively corners. We find that this occurs for $p>\log(N)/\log(2)$, as initially claimed. 
 
\subsection{Positive definite affinity matrices $J_{ij}$}\label{app:posdef}
We show that, at zero temperature, liquid mixtures with a positive definite affinity matrix $J_{ij}$ result in localization. To see this, we expand the matrix $J_{ij}$ in its eigenvectors, which gives
\begin{align}
\omega(\vec{\rho},0)= -\sum^N_{i,j=1}J_{ij}\rho_i\rho_j=-\sum_{\lambda=1}^Nc_{\lambda}\sum_{ij}v_i^\lambda v_j^\lambda\rho_i\rho_j=-\sum_{\lambda=1}^Nc_{\lambda}\left(\sum_{i}v_i^\lambda \rho_i\right)^2
\end{align}
where for now we assume the affinity matrix is full-rank, and  we have denoted its eigenvalues by $c_\lambda$ and  the corresponding eigenvectors  by $\vec{v}^\lambda$, with the later normalized to one, $|\vec{v}^\lambda|=1$.   By assumption, $c_{\lambda}>0$, as the affinity matrix is positive definite.  We now define the vectors $\vec{u}^\lambda=\vec{v}^\lambda/v^\lambda_{\rm max}$, where $v^\lambda_{\rm max}$ is the maximum (non-signed) component of $\vec{v}^\lambda$. We then have 
\begin{align}
\omega(\vec{\rho},0)=-\sum_{\lambda=1}^Nc_{\lambda}(v^\lambda_{\rm max})^2\left(\sum^N_{i=1}u_i^\lambda \rho_i\right)^2=-\sum_{\lambda=1}^Nb_\lambda m_\lambda^2
\end{align}
where we have defined the coefficients $b_{\lambda}=c_{\lambda}(v^\lambda_{\rm max})^2$ and the overlaps $m_\lambda=\sum^N_{i=1}u_i^\lambda \rho_i$, where $m_\lambda\in[-1,1]$. We can further define the coefficients $q_\lambda=m_\lambda^2/(\sum_\lambda m_\lambda^2)$, which are normalized $\sum^N_{\lambda=1} q_\lambda=1$. Using that $\sum^N_{\lambda=1} m_\lambda^2=||\vec{\rho}||_2^2$, we then have
\begin{align}
\omega(\vec{\rho})=-||\vec{\rho}||_2^2\sum_{\lambda=1}^Nb_\lambda q_\lambda\ge-{\rho}^2\sum_{\lambda=1}^Nb_\lambda q_\lambda\quad,
\end{align}
where in the last inequality we have used the Cauchy-Schwarz inequality.  As $b_\lambda>0$, due to the positive definiteness of the affinity matrix $J_{ij}$, the right-hand-side of the inequality is minimized for $\rho=1$ and $q_{\lambda}= \delta_{\lambda^\ast,\lambda}$ where $\lambda^\ast$ is the index with the  largest $b_{\lambda^\ast}$ coefficient. This corresponds to a localized state where all the density is on the component $i$ for which $v_i^\lambda$ is maximal, i.e. $|v_i^\lambda|=v^\lambda_{\rm max}$. Therefore, the right-hand-side is minimized for a localized state. Since for this localized state we actually have that $||\vec{\rho}||_2=\rho$, the inequality is saturated. Therefore, the localized state minimizes the grand-potential, which completes our proof. 

In summary, we have shown that for  positive semi-definite affinity matrices the low-temperature limit of a multi-component mixture corresponds to a localized state.  

\section{Limit of sparse targets $q\rightarrow 0$}
\label{app:sparse}
We discuss  chemical and mechanical equilibrium conditions for  retrieval states in the limit of  $q\approx 0$ and  for $v_3\approx 0$.    The main finding of this Section is that also in the sparse limit of $q=0$ a  finite $v_3>0$ is required to stabilize retrieval states.     An analogous analysis applies for $q\approx 1$.  

\subsection{Chemical equilibrium}

To take the limits $q\rightarrow 0$ and $v_3\rightarrow 0$, 
we make the substitution 
\begin{equation}
\hat{a}_\ast = \frac{a_\ast}{ \sqrt{q}}, \label{eq:tildeAst}
\end{equation}
in  Eqs.~(\ref{eq:chemEqQ_MM}) and (\ref{eq:chemEqRhoQ_MM}), 
and consider the expansions 
\begin{equation}
\hat{a}_\ast = \hat{a}^{(0)}_\ast + v_3 \hat{a}^{(1)}_\ast + O(v^2_3)\label{eq:linearOrderPerturbA}
\end{equation}
and 
\begin{equation}
\rho_\ast = \rho^{(0)}_\ast + v_3 \rho^{(1)}_\ast + O(v^2_3 ) .
\end{equation}

Solving the resultant equations towards the densities we get  the coefficients
 \begin{equation}
\rho^{(0)}_\ast  = \frac{1}{1+e^{-\mu_N}}\label{eq:rhoAst1}
\end{equation} 
and 
\begin{equation}
\rho^{(1)}_\ast  =  - \frac{e^{-\mu_N}}{2(1+e^{-\mu_N})^4}.\label{eq:rhoAst2}
\end{equation}
Note that in the limit of sparse target states, the densities do not depend on the overlap $\hat{a}_\ast$, and hence retrieval regions are enriched in a few components, but their overall density will not differ from the exterior density.   Also, note that the correction term $\rho^{(1)}_\ast$ is negative as $v_3$ quantifies the strength of a repulsive potential.  

For the coefficients of the overlap parameter, we obtain the implicit equation
\begin{eqnarray}
  \hat{a}^{(0)}_{\ast} &=&  {\rm exp}\left(  \hat{a}^{(0)}_{\ast}  \hat{v}_2\right)  -1 , \label{eq:Atilde1}
\end{eqnarray}
where  $\hat{v}_2 = v_2/(1+e^{-\mu_N})$.
Apart from the trivial solution $\hat{a}^{(0)}_\ast = 0$, Eq.~(\ref{eq:Atilde1}) admits the solution  
\begin{equation}
  \hat{a}^{(0)}_\ast = \left\{\begin{array}{ccc}-1 - \frac{1}{\hat{v}_2}W_{-1}\left(-e^{-\hat{v}_2}\hat{v}_2\right)&{\rm if}& \hat{v}_2\leq 1,\\ -1 - \frac{1}{\hat{v}_2}W_{0}\left(-e^{-\hat{v}_2}\hat{v}_2\right)&{\rm if}& \hat{v}_2> 1, \end{array}\right. \label{eq:atilde}
\end{equation}
where $W_{0}(x)$  is the principal branch  of the Lambert-W function  that solves $x =W e^{W}$, and  $W_{-1}(x)$   is the second branch of the Lambert-W function.     The coefficient for the linear term  in perturbation theory is given by 
\begin{eqnarray}
\hat{a}^{(1)}_\ast  = \hat{a}^{(0)}\frac{ e^{-\mu_N+\hat{a}^{(0)}_\ast\hat{v}_2} }{2 (1+e^{-\mu_N})^{4}} \frac{4+2\hat{a}^{(0)}_\ast+v_2+2(2+\hat{a}^{(0)}_\ast)\cosh(\mu_N)}{(-1 + \hat{v}_2 e^{\hat{a}^{(0)_\ast}\hat{v}_2})}. \label{eq:atilde1}
\end{eqnarray} 
Note that when $ \hat{a}^{(0)}_\ast=0$, $\hat{a}^{(1)}_\ast$ is in general not zero-valued, as the denominator also converges to zero in this limit, yielding a finite value for $\hat{a}^{(1)}_\ast$.

In Fig.~\ref{fig:aTotalLinear} we plot    $\hat{a}_\ast$ as a function of $\hat{v}_2$ up to linear order in perturbation theory.    Observe that $\hat{a}^{(0)}_\ast$ is positive for $\hat{v}_2<1$ and negative for $\hat{v}_2>1$, while for $v_3>0$ the transition point where $\hat{a}_\ast$ changes sign is larger than one.    Interestingly, for  $\hat{v}_2\rightarrow 0$  the overlap $\hat{a}_\ast$ diverges.

\subsection{Mechanical equilibrium}
Having determined the chemically stable retrieval states in the limit of sparse target vectors, we now determine their mechanically stable branches.   To this aim, we expand the coefficients $c_1$ and $c_2$ that appear in the stability conditions (\ref{eq:1Q}) and (\ref{eq:2Q}) in $v_3$,  viz.,
\begin{equation}
c_1= c^{(0)}_1 + v_3 c^{(1)}_1 +O(v^2_3) \label{eq:c1Linear}
 \end{equation}
 and
\begin{equation}
c_2 = c^{(0)}_2 + v_3 c^{(1)}_2   + O(v^2_3)\label{eq:c2Linear}
 \end{equation}
 and we determine the above four coefficients from a Taylor expansion in $v_3$ of the  Eqs.~(\ref{eq:c1})  and (\ref{eq:c2})  in the limit $q\rightarrow 0$.   

For the zero-order coefficients, that determine stability at $v_3=0$,  we get 
 \begin{equation}
 c^{(0)}_1 = 1+e^{-\mu_N} \label{eq:c01}
 \end{equation}
and 
\begin{equation}
c^{(0)}_2 = \frac{\hat{a}^{(0)}}{1+\hat{a}^{(0)}}(1+e^{-\mu_N}).   \label{eq:c02}
 \end{equation}
Substitution of  Eqs.~(\ref{eq:c01}) and (\ref{eq:c02}) into Eqs.~(\ref{eq:1Q}) and (\ref{eq:2Q}), we find that for all values of $\tilde{v}_2$   retrieval states are {\it not} mechanically stable  at $v_3=0$.  Indeed,   
for $\hat{v}_2\leq 1$ it holds that $\hat{a}^{(0)}>0$, and therefore $c^{(0)}_2>0$.  Consequently,   the inequality (\ref{eq:2Q})  determines mechanical stability,  which reads here  $\hat{a}^{(0)}_{\ast}  \leq  (1-\hat{v}_2)/\hat{v}_2 $, and is not satisfied for $\hat{v}_2\leq 1$.   Analogously, for $\hat{v}_2\geq 1$ it holds that $\hat{a}^{(0)}<0$, and therefore $c^{(0)}_2<0$.  Consequently, we require the  inequality (\ref{eq:1Q}) that reads $\hat{v}_2\leq 1$ in the limit $q\rightarrow 0$.   

\begin{figure}[ht!]
\centering
\includegraphics[width=0.4\textwidth]{ 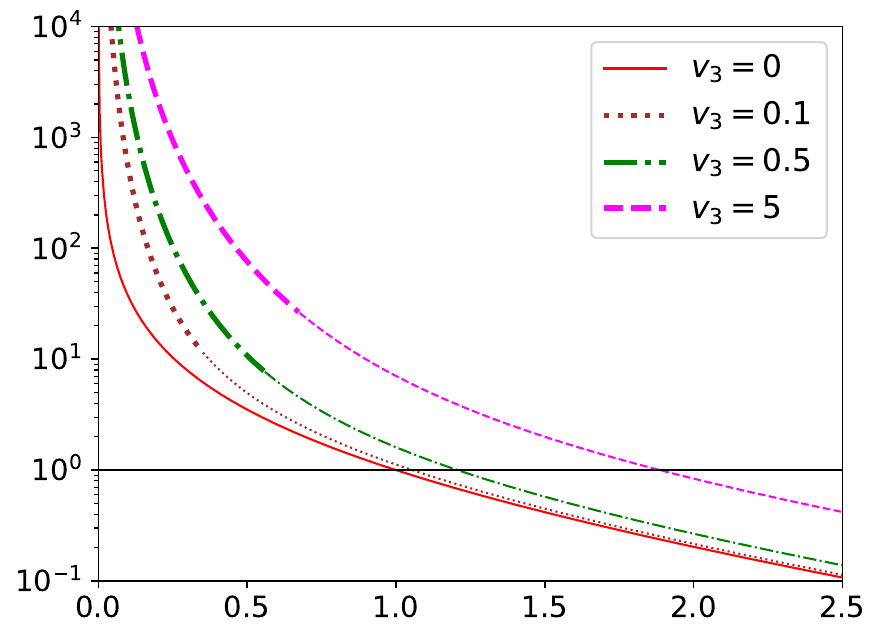}
 \put(-225,70){$\hat{a}_\ast+1$}
  \put(-95,-10){$\hat{v}_2$}
    \caption{Plot of the chemically stable value of the overlap $\hat{a}_\ast = a_\ast /\sqrt{q}$ of retrieval states at $q=0$ as a function of $\hat{v}_2 = v_2/(1+\exp(-\mu_N))$  for small values of $v_3$.    We used the   perturbation theory up to linear order, see (\ref{eq:linearOrderPerturbA}),  with coefficients determined by  the formulae (\ref{eq:atilde}) and  (\ref{eq:atilde1}).  We have set $\mu_N=1$ and $v_3$ is given as in the legends.   The thick lines and thin lines denote the mechanically stable  and unstable branches, respectively.   For $v_3=0$ the retrieval state  is unstable for all values of $\hat{v}_2$.    The solid horizontal line denotes $\hat{a}_\ast = 0$ and is a guide to the eye.          } \label{fig:aTotalLinear}
\end{figure}

Taken together, retrieval states are not stable for $v_3=0$, and we require a nonzero $v_3$ to stabilise them.   The linear-order coefficients in Eqs.~(\ref{eq:c01}) and (\ref{eq:c02}) read 
\begin{equation}
c^{(1)}_1 = \frac{2+3 e^{-\mu_N}}{2(1+e^{-\mu_N})^2} \label{eq:c11}
\end{equation}
and 
\begin{eqnarray}
{c^{(1)}_2}= \frac{e^{-\mu_N}}{2\left(1+\hat{a}^{(0)}\right)^2\left(1+e^{\mu_N}\right)^2} \left(2\hat{a}^{(1)}\left(1+e^{\mu_N}\right)^3
-\hat{a}^{(0)}(1+\hat{a}^{(0)})e^{2\mu_N}\left\{1+2\hat{a}^{(0)}+2(1+\hat{a}^{(0)})e^{\mu_N}\right\}\right).\nonumber\\
\label{eq:c12}
 \end{eqnarray}
Substitution of (\ref{eq:c01}) and (\ref{eq:c11}) in (\ref{eq:c1}), and substitution of (\ref{eq:c02}) and (\ref{eq:c12}) in (\ref{eq:c2}), yields  expressions for $c_1$ and $c_2$ up to linear order in $v_3$, which we can substitute in Eqs.~(\ref{eq:1Q}) and (\ref{eq:2Q}) to determine the mechanically stability of retrieval states up to linear order in perturbation theory.       In Fig.~\ref{fig:aTotalLinear}, we depict the stable branches of the retrieval state with thick lines, demonstrating that  stable retrieval is possible  at small values of $v_2$ when $v_3>0$.

\begin{figure}[ht!]
\centering
\includegraphics[width=0.4\textwidth]{ 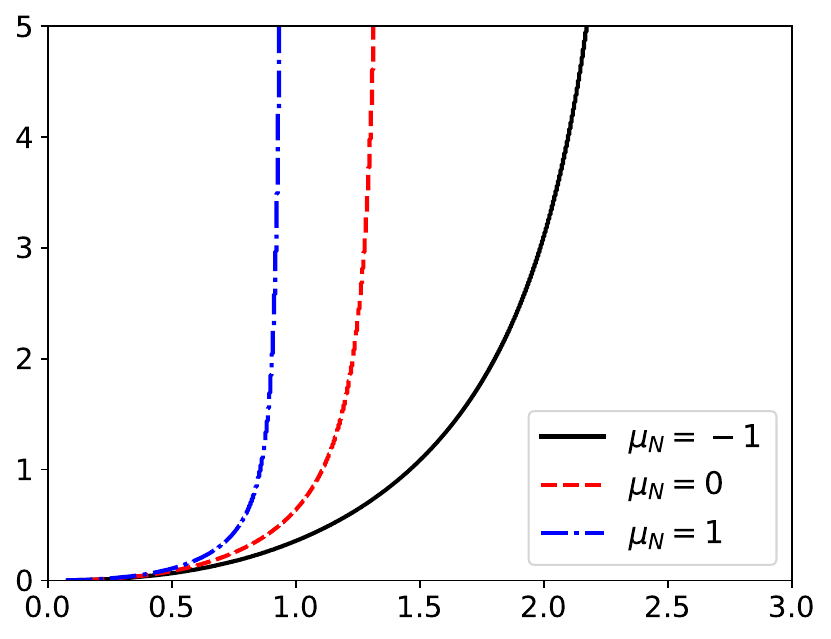}
 \put(-220,70){\large$v_3$}
  \put(-110,-10){\large$v_2$}
    \caption{Lines of marginal stability,   separating a stable retrieval state at small values of $v_2$ from an unstable retrieval state at large values of $v_2$, calculated up to linear order in perturbation theory ($v_3$ being the small parameter) and for $q=0$.   Lines are obtained by using the linear expressions (\ref{eq:c1Linear}) and (\ref{eq:c2Linear}) for $c_1$ and $c_2$ in the equality of Eq.~(\ref{eq:1Q})  [we only find stability in the region with $c_2>0$, and thus (\ref{eq:2Q}) is no longer relevant] evaluated at  $q=0$.       } \label{fig:phaseDiagramq0}
\end{figure}

In Fig.~\ref{fig:phaseDiagramq0} we plot the lines of marginal stability up to linear order in $v_3$.   Above these lines, the retrieval state is stable.   Importantly, for finite values of $v_2$ we require a finite $v_3$ to render  retrieval states mechanically stable.   Note that according to Fig.~\ref{fig:phaseDiagramq0}  stable retrieval states are not possible for large enough values of $v_2$, as an infinitely large $v_3$ is required to stabilise states beyond a finite value of $v_2$.   However, it should be stressed that this is a perturbative result up to linear orders in $v_3$, and we expect this effect to disappear when considering higher orders in perturbation theory.

\section{Numerical methods}
\label{app:methods}
We elaborate on the numerical methods used to generate Figures~\ref{fig:scaling} and \ref{fig:loc}.  

The numerical results in these figures are obtained by numerically minimising the  grand-potential functional $\omega(\vec{\rho}.\mu^{\rm res})$ with the  truncated Newton method given an initial state $\vec{\rho} = \vec{\rho}^{\rm init}$.  Specifically, the \textit{TNC} algorithm of the \textit{Scipy Optimization library} was used. 

The key observable in  Fig.~\ref{fig:scaling} is the size of the basin of attraction $\mathcal{S}$ which is a function of the $p$ selected hypersymmetric target states and a sequence $\vec{\rho}^{\rm init}(r)$ of initial states, where $r=m/j$ and  $j\in \left\{1,2,\ldots, m\right\}$; in the Fig.~\ref{fig:scaling}   $m=50$.  The size of the basin of attraction $\mathcal{S}$  is the  maximal value of $r$ for which the minimization algorithm converges to the target state (we always consider enough iterations for the minimizer to find a minimum).

We specify how the $m$ initial states $\vec{\rho}^{\rm init}(r)$  are constructed.   The component densities of the initial states are set to either of three values: a chosen number $Nr$ of the densities is set to the value $\rho_\ast$ in the homogeneous state given by Eq.~\eqref{eq:crit_point_homogeneous}; the remaining $N(1-r)$ densities are set to the pair of values given by the target state $\vec{\rho}^\alpha_\ast$ with component densities $\rho_i^\alpha$ as defined in Eq.~\eqref{eq:ansatz}, and where $a=a_\ast$ and $\rho=\rho_\ast$ from Eqs.~\eqref{eq:rho_a_minA} and \eqref{eq:rho_a_minB}.  The $Nr$ components with densities $\rho_\ast$  are selected  uniformly at random amongst the $N$ possible component.   Moreover,  these components are selected independently for different values of $r$. 

Next, we specify how $\mathcal{S}$ is determined from a given sequence of initial states $\vec{\rho}^{\rm init}(r)$.    For this, we determine the maximal value of   $r$ for which the overlap
\begin{equation}
a(\vec{\rho}^\alpha_\ast, \vec{\rho}^{\rm final})  = \frac{\sum^N_{i=1}\rho^\alpha_{\ast,i}\rho^{\rm final}_i}{\sum^N_{i=1}\rho^{\rm final}_i}
\end{equation}
between the target state $\vec{\rho}^\alpha_\ast$ and the final converged result $\vec{\rho}^{\rm final}(r)$ of the minimising algorithm is larger or equal than $0.99$, i.e., 
\begin{eqnarray}
{\mathcal{S}(\vec{\gamma}^{\alpha},\left\{\vec{\gamma}^\beta\right\}_{\beta=1,2,\ldots,p}; \left\{\vec{\rho}^{\rm init}(r)\right\}_{r=1,2,\ldots,m})} =  \frac{{\rm max}\left\{r\in \left\{1,\ldots,m\right\}: a(\vec{\rho}^\alpha_\ast, \vec{\rho}^{\rm final}(r)) \geq  0.99\right\}}{m}.\nonumber 
\end{eqnarray} 
Lastly we specify the meaning of the average value $\langle \mathcal{S}\rangle$.   The average is taken over both the random choice of the $p$ hypersymmetric targets, as described in Sec.~\ref{app:symmetric}, the random choice of the selected target state $\vec{\gamma}^{\alpha}$, and the random choice of the  components of the sequence $\vec{\rho}_{\rm init}(r)$ that are set equal to $\rho_\ast$.   In Fig.~\ref{fig:scaling} the average $\langle \mathcal{S}\rangle$ is estimated as an empirical mean over $20$ realisations of both the targets and the choice of components that are set equal to $\rho_\ast$.   Error bars denote the standard deviation on the empirical mean.

For Fig.\ref{fig:loc}, the initial states $\vec{\rho}^{\rm init}$ for the truncated Newton method applied to $\omega$ are generated from the, so-called, flat Dirichlet distribution, which corresponds with a uniform sampling over the the standard simplex.  Figure~\ref{fig:loc}A is constructed from  $10^6$ initial configurations, and each point of Fig.\ref{fig:loc}B is an empirical mean over $10^5$ initial configuration points.

\section{Scaling analysis of Ref. \cite{jacobs2021self} for finite $q$}\label{app:jacobscale}
As stated in the main text, one key difference between this paper and  Ref. \cite{jacobs2021self} is that here we consider the case of $q$ is finite. That is, we showed that for finite $q$ the number of targets that can be simultaneously stable is $p=N-1$, provided a cubic non-linearity is present. While Ref.~\cite{jacobs2021self} does not include a cubic term, we now show, following the scaling analysis in \cite{jacobs2021self}, that the retrieval works at finite $q$ for  a non-extensive number of patterns $p$, and hence the capacity $p/N$ converges to zero for large $N$.   In other words, to obtain a nonzero retrieval capacity, Ref.~\cite{jacobs2021self} requires $q = 0$.

We start by considering the following expression
\begin{align}
p(N-Q)\left(\frac{pQ^2}{N^2}\right)^Q\sim 1,
\end{align}
which corresponds to the condition identified in Ref. \cite{jacobs2021self} for successful retrieval (see Supplementary Information section C in Ref. \cite{jacobs2021self}). Since we aim to investigate the extensive regime, we define the sparsity parameter, $q =Q/N$, so that
\begin{align}
p^{qN+1} \sim \frac{1}{N(1-q)q^{2qN}}.
\end{align}
Taking logarithms this can be written as 
\begin{align}
\log p \sim -\frac{1}{qN+1}\log N -\frac{1}{q^{N+1}}\log(1-q) - \frac{2qN}{qN+1}\log q.
\end{align}
To recover the result in Ref.~\cite{jacobs2021self} (Eq. C3 in the Supplementary Information) one must take the following limit: $q\to0$ and $N\to\infty$, with  $qN=Q$ fixed. Instead, to consider this Paper's limit, we keep $q$ finite while taking $N\to\infty$. The result is  
\begin{align}
p \sim \frac{1}{q^2}.
\end{align}
This result shows that the number of targets $p$ that can be reliably retrieved is constant as $N$ increases. In other words, in this regime $p/N$ converges to 0. This is in agreement with the result of the present Paper that retrieval is not possible in absence of higher order interactions when $Q=qN$ with $q$ finite. A different way of seeing this is that in Ref.~\cite{jacobs2021self} the targets are localized, and therefore there is no difference between localization and retrieval, and no need for a cubic term. This is also in agreement with Fig.~\ref{fig:sparse}, which shows that as targets become more sparse cubic nonlinearities become less important.

\section{Alternative choices of non-linear interactions}
\label{app:alt-nonlin}

\begin{figure}[h]
\centerline{\includegraphics[width = 12cm]{ 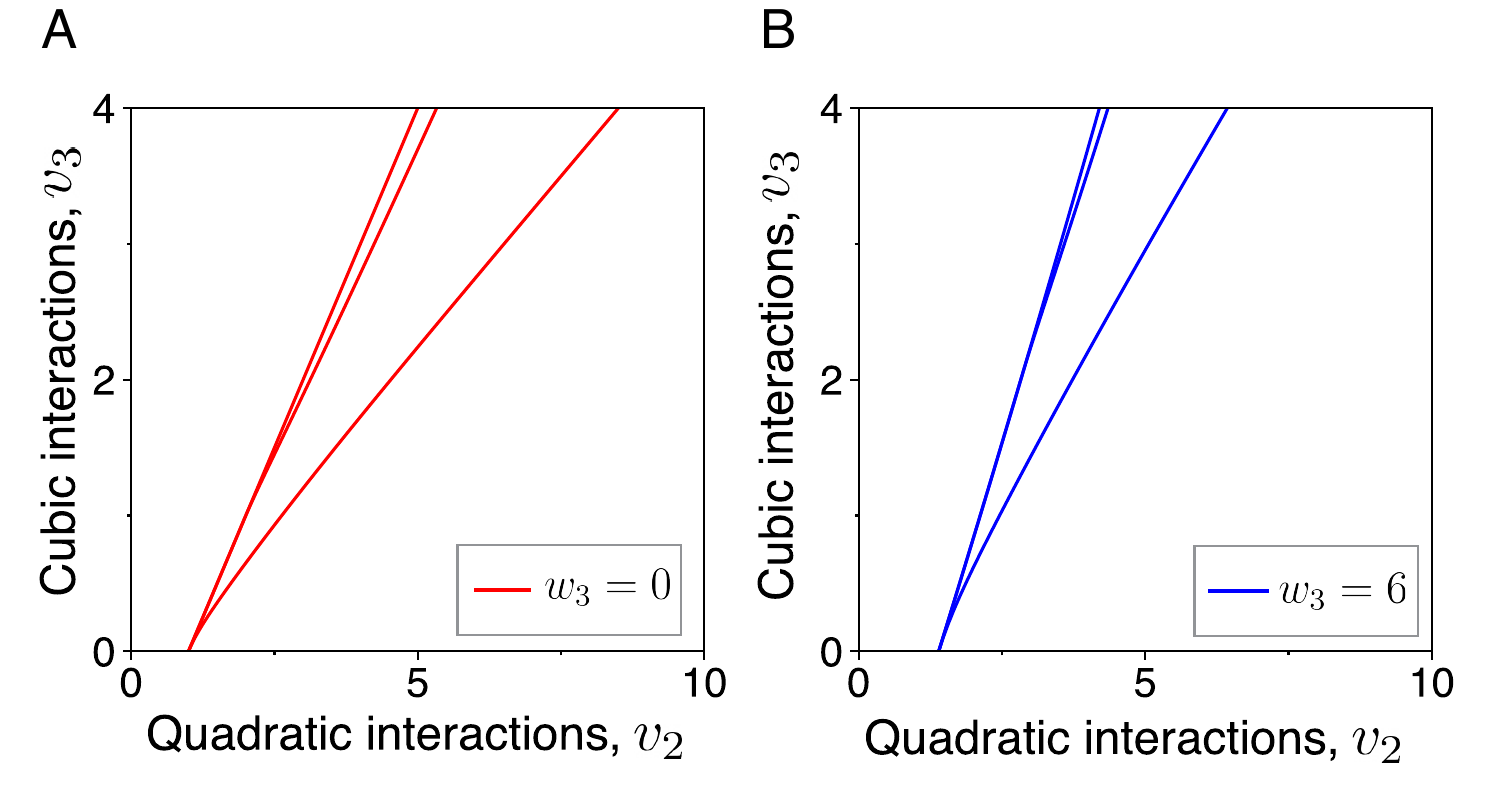}}
\caption{\label{fig:const_rep}{ \it Stability diagrams for  $w_3=0$ (Panel A) and $w_3=6$ (Panel B).}    Spinodal lines delimit the regions where either the  homogeneous state or all retrieval states are stable as a function of $v_2$ and $v_3$ with  $\mu_N=10$ (similar to Fig.\ref{fig:phase_diag} in the main text).  The plotted lines are obtained from  the equalities in Eqs.~\ref{eq:crit_temp_homogeneous}, \ref{eq:stability_ansatz} and \ref{eq:stability_ansatz2} from the manuscript using $\rho_*$ given by Eq.~\eqref{eq:chemEqRhoQ_MM}.}
\end{figure}

In the main text, we considered three body self-repulsive interactions. This choice of interaction was motivated by its simplicity. We now show that including repulsion among different species of components only affects the main results of this paper  quantitatively. To show this, we add to the internal energy Eq.~\ref{eq:u} the following term:
\begin{equation}\label{eq:global_rep}
+w_3\sum_{i,j,k}\rho_i\rho_j\rho_k,
\end{equation}  
where $w_3\geq 0$ measures the strength of the repulsion among different types of particles. Following the same approach as in the main text, the conditions of chemical equilibrium for this modified model can be derived. The resulting equation for $a_*$ remains the same as in the case for $w_3=0$, while the equation for $\rho_*$ is modified. Concerning mechanical equilibrium, since the repulsive term in Eq.~\eqref{eq:global_rep} is constant and positive, the formal equations  for mechanical stability do not change explicitly, but nevertheless, the lines of marginal stability are altered with $w_3$ as the expression of $\rho_*$ depends on it.  In Fig.~\ref{fig:const_rep} we show stability diagrams in the presence (and absence) of the global repulsive term $w_3$. We  conclude that the $w_3$ does not change qualitatively the paper's main results, but quantitatively it does decrease the stable retrieval region.

\begin{figure}[h]
\centerline{\includegraphics[width = 12cm]{ 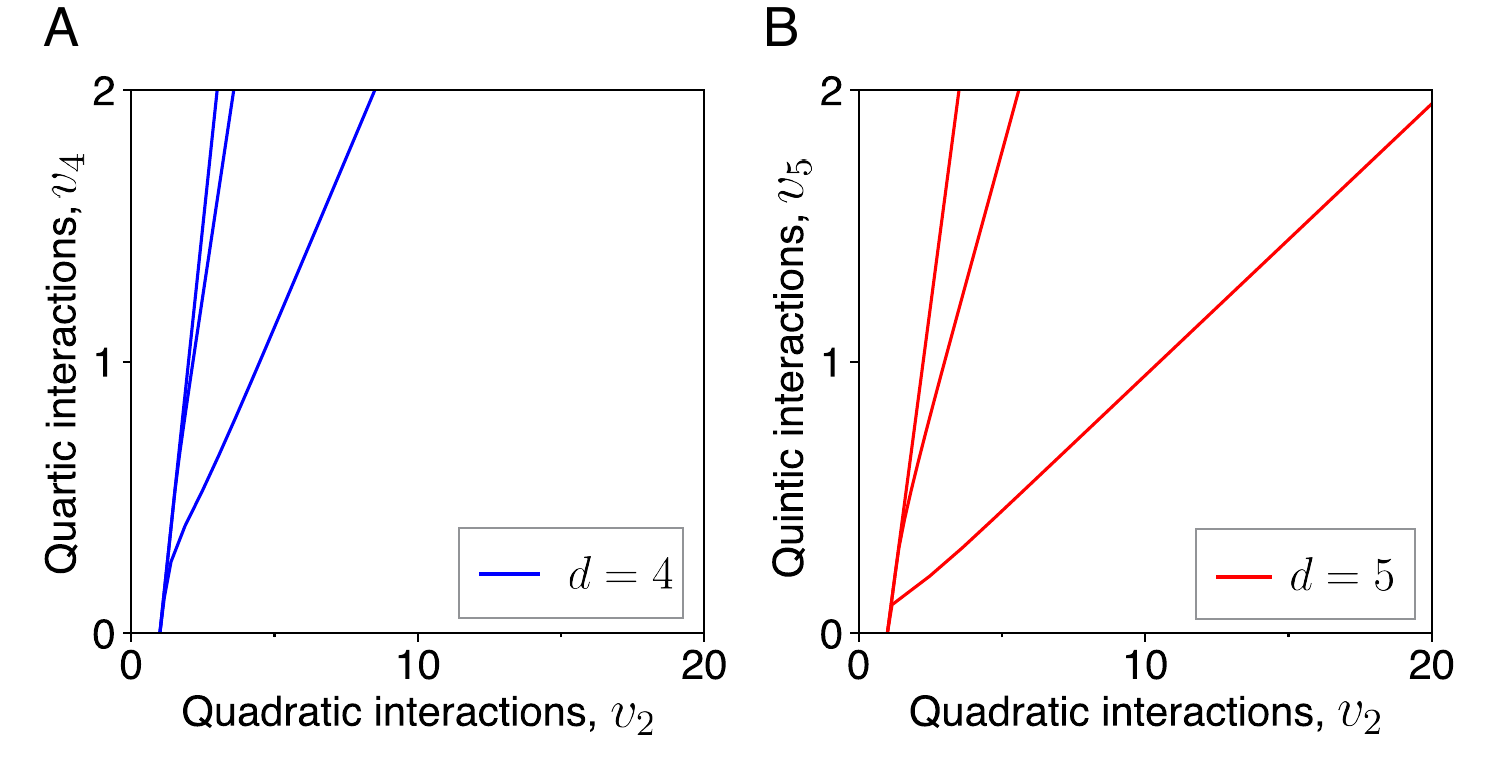}}
\caption{\label{fig:stab_high} \textit{ Stability diagrams for interaction matrices with a nonlinear repulsion term  of order $d=4$ (Panel A) or $d=5$ (Panel B).} 
These figures are analogous to the $d=3$ diagram in Panel A of Figure \ref{fig:phase_diag} of the main text, but the cubic interaction is replaced with the term in Eq.~(\ref{eq:b0}).   The lines delimit the regions where either the   homogeneous state (left line) or all retrieval states (the other two lines) are stable for parameter value $\mu_N=1.8$. }
\end{figure}

To further illustrate that the results of this Paper are not specific to the chosen model, we  consider non-linear terms with interactions that are of  order higher than cubic. Also in this case the paper's main results are qualitatively the same. Specifically, we add to the interaction matrix  the following general non-linearity:
\begin{equation}\label{eq:b0}
v_d\frac{ \sum_i\rho_i^d N^{d-1}}{d(d-1)},  
\end{equation}  
where $v_d$ measures the strength of the interaction. Following the same derivation as before, we  can generalize the conditions for chemical and mechanical equilibrium (equations not shown). In Fig.\ref{fig:stab_high}A and B we plot the stability diagram for $d = 4$ and $d=5$, respectively. We find a similar phenomenology as the one studied in detail in the paper $(d = 3)$.   Interestingly, we find that the retrieval region becomes larger with increasing order of the non-linearity.

Furthermore, we investigate the robustness of the retrieval capabilties of the liquid Hopfield model towards variability in the strength of the  self-repulsive interactions.    To this purpose, we re-define the cubic interactions in the following way:
\begin{equation}
    \frac{1}{6}\sum_{i=1}^{N}\tilde{v}_{3i}\rho_i^3N^{2},
\end{equation}
where $\tilde{v}_{3i} = v_{3} + g_{i}$ is the strength of the perturbed cubic interaction and the  $g_i$ are independent and identically distributed random variables drawn from   a normal (Gaussian) distribution with mean $0$ with standard deviation $\sigma$.

\begin{figure}[h]
\centerline{\includegraphics[width = 12cm]{ 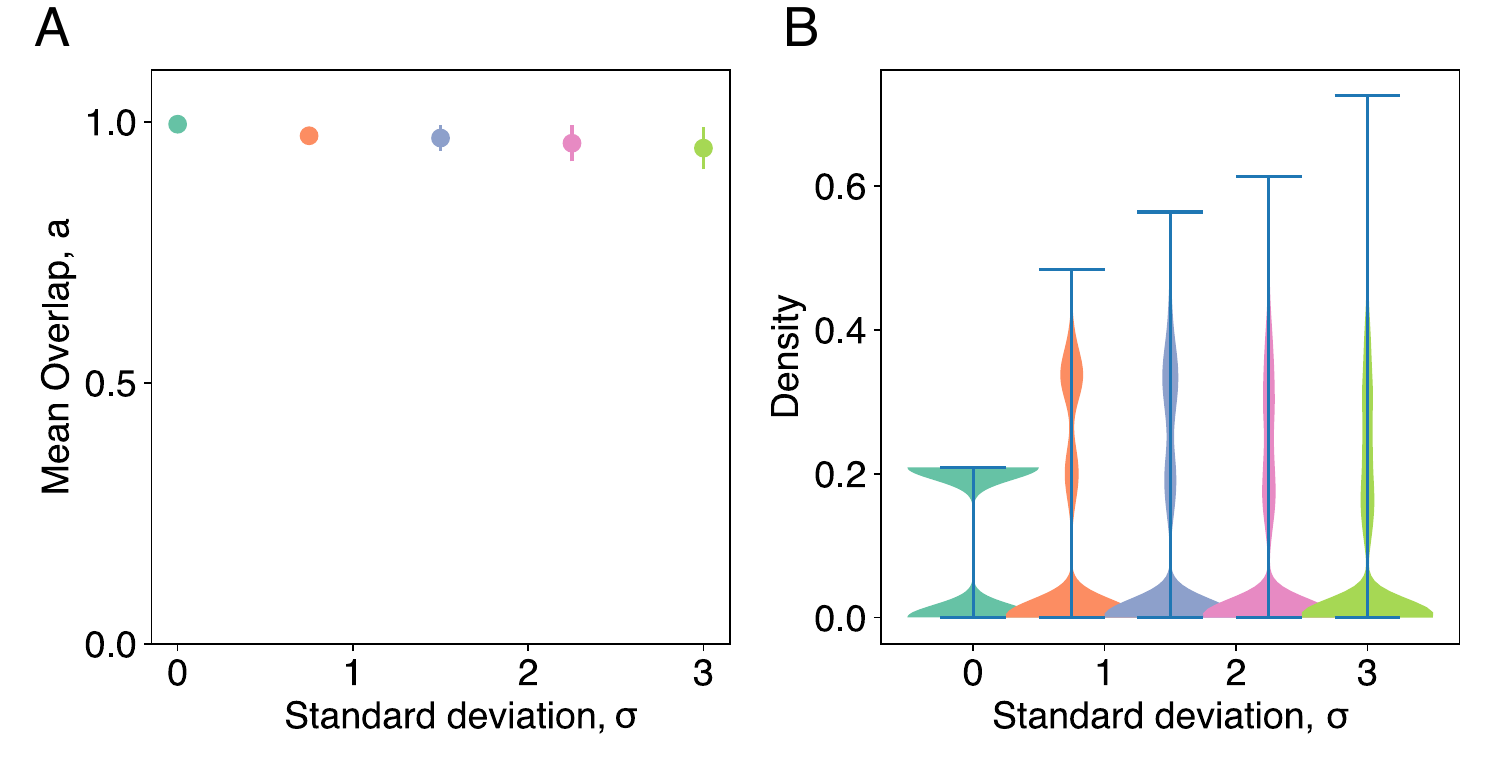}}
\caption{\label{fig:cubic_perturb}{{\it The effect of perturbing the cubic interactions.}  {\bf A. } Mean (normalized) overlap, $a$, as a function of the standard deviation, $\sigma$ for $N = 8$, $v_2 = 7 $, $v_3 = 4$ and $\mu_N=0$.  {\bf B. } Distributions of the density states $\rho_i$ as a function of the standard deviation, $\sigma$. The range of the densities of retrieval states get spread out as $\sigma$ increases.  A total of $1000$ repetitions were performed for each value of $\sigma$. }}
\end{figure}

We numerically minimize the grand-potential in the presence of this perturbation and study the minima. 
As demonstrated in Fig.\ref{fig:cubic_perturb}, retrieval  works in the liquid Hopfield model even with fluctuating cubic interactions. For instance, Fig.~\ref{fig:cubic_perturb}A shows that the overlap parameter $a$ remains close to $1$, even when the fluctuations in the  cubic term, quantified by $\sigma$, are large. Thus, even in this case the free energy minima show enrichment and depletion patterns that match the encoded targets. In addition, in Fig. \ref{fig:cubic_perturb}B we show that for variable cubic repulsive interactions    the component densities $\rho_i$ span a large range.

Overall, these results show that the liquid Hopfield model, as defined in this Paper, is a minimal working example.    Adding  across-species interactions and more complex non-linear terms to the original model yields a liquid mixture with similar retrieval capabilities.  In addition, these model variations also exhibit a  trade-off between retrieval and localization. 

\color{black}

\section{Robustness analysis of the Liquid Hopfield Model - chemical potential perturbations}\label{app:chem_pert}

In the main paper we  have studied liquid mixtures for which the chemical reservoir of all  components have the same chemical potential $\mu_i=\mu^{\rm res}$.    Here,  we study the effect of variability in the chemical potentials on the retrieval capabilities of the liquid Hopfield model.  In particular, we consider the liquid Hopfield model with chemical potentials 
\begin{equation}
\mu_i = \mu^{\rm res} + g_i\quad,
\end{equation}
where  the $g_i$ are independent and identically distributed random variables a random variable drawn from a normal (Gaussian) distribution with mean  $0$ and standard deviation $\sigma$.  

\begin{figure}[h]
\centerline{\includegraphics[width = 16cm]{ 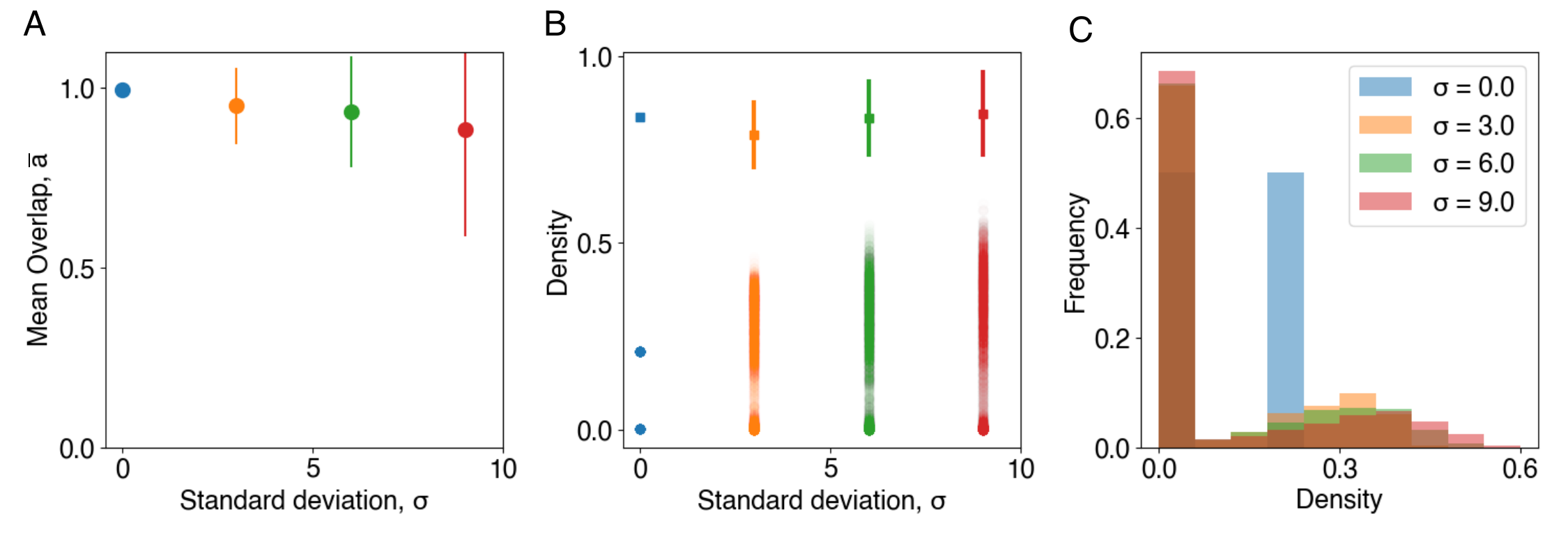}}
\caption{\label{fig:noise}{ {\it The effect of perturbing the chemical potentials.}  {\bf A. } Mean (normalized) overlap, $a$, as a function of the standard deviation, $\sigma$ for $N = 8$, $v_2 = 12 $, $v_3 = 4$ and $\mu_N=0$.  {\bf B. }Total density $\rho_\ast$ of the retrieval state (squares) and the corresponding densities of enriched  and depleted components (circles)  as a function of the standard deviation, $\sigma$. The range of the densities of retrieval states get spread out as $\sigma$ increases. {\bf C. } Histograms showing the frequency of $\rho_i$ for different values of $\sigma$.   A total of $1000$ repetitions were performed for each value of $\sigma$. }}
\end{figure}

We  numerically minimise the   grand potential of the liquid Hopfield model with disordered chemical potential to determine its retrieval properties.
Figure~\ref{fig:noise} demonstrates that retrieval also works in the liquid Hopfield model with random  chemical potentials, albeit now with high variability in the densities of the individual components. Indeed, Fig.~\ref{fig:noise}A shows that the overlap parameter, $a$, remains close to $1$, even for large values of  $\sigma$, the standard deviation of the chemical potentials. In other words, the minima of the free energy exhibit patterns of enrichment and depletion consistent with the encoded targets, showing the success of retrieval even in the  case with variable chemical potentials. The variability of the chemical potentials yields stable states that can have large variability in the densities $\rho_i$ of the individual components, as we show in  
Figure~\ref{fig:noise}B and \ref{fig:loc}C. 
Therefore, the retrieval capabilities of the liquid Hopfield model are   robust to variabilities in concentrations spanning a wide range.

\end{document}